\documentclass[11pt]{article}
\usepackage{amsmath,amssymb,epsf,color,slashed}
\usepackage{mathrsfs}

\def\ft#1#2{{\textstyle{\frac{\scriptstyle #1}{\scriptstyle #2} } }}
\def\fft#1#2{{\frac{#1}{#2}}}
\def\cP{{{\cal P}}}

\def\pa{\slashed{\partial}}

\def\be{\begin{equation}}
\def\ee{\end{equation}}
\def\ba{\begin{array}}
\def\ea{\end{array}}
\def\bea{\begin{eqnarray}}
\def\eea{\end{eqnarray}}
\def\bd{\begin{displaymath}}
\def\ed{\end{displaymath}}

\def\nn{\nonumber}

\def\del{\partial}
\newcommand{\w}[1]{\\[0.#1cm]}
\newcommand{\eq}[1]{(\ref{#1})}

\newcommand{\opsi}{\bar{\psi}}

\newcommand\T{\rule{0pt}{2.6ex}}

\newsavebox{\uuunit}
\sbox{\uuunit}
    {\setlength{\unitlength}{0.825em}
     \begin{picture}(0.6,0.7)
        \thinlines
        \put(0,0){\line(1,0){0.5}}
        \put(0.15,0){\line(0,1){0.7}}
        \put(0.35,0){\line(0,1){0.8}}
       \multiput(0.3,0.8)(-0.04,-0.02){12}{\rule{0.5pt}{0.5pt}}
     \end {picture}}

\def\cell{c_\ell}
\makeatletter
\@addtoreset{equation}{section}
\makeatother

\def\1{{\cal O}_1}
\def\2{{\cal O}_2}
\def\T2{\widetilde{\cal O}_2}
\def\3{{\cal O}_3}
\def\4{{\cal O}_4}

\def\im{{i}}
\def\ep{\epsilon}

\begin{document}

\begin{flushright}
\hfill{ \
MIFPA-12-12\ \ \ \ }
\end{flushright}

\vspace{25pt}

\begin{center}

{\large \bf Spectrum of Higher Derivative $6D$ Chiral Supergravity}

{\large\bf on Minkowski $\times S^2$}

\vspace{15pt}

Y. Pang, C.N. Pope and E. Sezgin

\vspace{10pt}

{\it George P. \& Cynthia Woods Mitchell  Institute
for Fundamental Physics and Astronomy,\\
Texas A\&M University, College Station, TX 77843, USA}

\vspace{40pt}

\underline{ABSTRACT}

\end{center}

\bigskip

Gauged off-shell Maxwell-Einstein supergravity in six dimensions with $N=(1,0)$ supersymmetry has a higher derivative extension afforded by a supersymmetrized Riemann squared term. This theory admits a supersymmetric Minkowski $\times S^2$ compactification with a $U(1)$ monopole of unit charge on $S^2$. We determine the full spectrum of the theory on this background. We also determine the spectrum on a non-supersymmetric version of this compactification in which the monopole charge is different from unity, and we find the peculiar feature that there are massless gravitini in a representation of the $S^2$ isometry group determined by the monopole charge.

\vspace{15pt}

\thispagestyle{empty}

\newpage

\tableofcontents

\newpage


\section{Introduction}


Higher-derivative supergravities are of considerable interest,
especially when they arise as low-energy effective actions of
string theories with higher-derivative corrections proportional to powers
of the slope parameter $\alpha'$. However, their construction is
notoriously difficult, in part due to the fact that supergravities exist
only on-shell in ten dimensions. In view of this difficulty, the
compactifications of these theories are rarely studied.
In order to gain insights into the compactification of higher-derivative
theories, it is instructive to investigate the issue in the simpler
situation of lower-dimensional supergravities with higher-derivative terms,
postponing for the present the question of how they may arise from
ten dimensions. An important technical advantage is that in some
lower-dimensional cases, off-shell formulations of the supergravity theories
exist. This leads us to consider in particular ${\cal N}=(1,0)$
supergravity in six dimensions, which is the highest dimension, and the
highest degree of supersymmetry, for which a supergravity with an
off-shell formulation is known. The off-shell formulation of this
supergravity was constructed in \cite{Bergshoeff:1985mz,Coomans:2011ih},
and a higher-derivative extension with an off-shell supersymmetrized
Riemann-squared term was obtained in \cite{Bergshoeff:1986wc,Bergshoeff:2011xn}.
The gauging of the $U(1)$ R-symmetry in the presence of this higher-derivative
extension has also recently been obtained \cite{Coomans:2012ax}.  The model
has two parameters, namely an overall coefficient $M^{-2}$ in front of the
higher-derivative superinvariant in the action, and the gauge-coupling
constant $g$.

   In the present paper, we shall study the six-dimensional gauged ${\cal N}=(1,0)$
theory with the Riemann-squared term constructed in \cite{Coomans:2012ax}.  In the
absence of the curvature-squared terms the model is an (off-shell) version
of the Salam-Sezgin theory constructed long ago \cite{Salam:1984cj}.  It was
shown in \cite{Salam:1984cj} that the model had the unusual feature of admitting
a supersymmetric Minkowski$_4\times S^2$ vacuum, in which there is a
$U(1)$ monopole flux with charge $q=\pm1$ on the $S^2$ internal space.
A remarkable feature of the theory with the Riemann-squared extension
is that the Minkowski$_4 \times S^2$ background continues to be a
supersymmetric solution \cite{Coomans:2012ax}.
It also admits non-supersymmetric Minkowski$_4\times S^2$
backgrounds in which the quantised monopole charge $q$ is larger than 1.

Our focus in this paper is to study the spectrum of the Kaluza-Klein
states in the fluctuations around the Minkowski$_4\times S^2$
background. As far
as we are aware, such a Kaluza-Klein spectral analysis of
a higher-derivative supergravity around a background with non-abelian
symmetries has not previously been carried out. Even in the
much simpler $S^2$ reduction of the Salam-Sezgin model discussed in
\cite{Salam:1984cj}, the situation is of considerable interest because of the
very unusual feature of obtaining non-abelian symmetries from a sphere
reduction, whilst obtaining a Poincar\'e rather than AdS supergravity in
the lower dimension. As
expected, the states assemble into ${\cal N}=1$ four-dimensional
supermultiplets.  In the model constructed in \cite{Coomans:2012ax} with
the higher-order Riemann-squared extension, we find a number of novel features
associated with the occurrence of higher-order
wave operators, and the fact that certain fields that were purely
auxiliary prior to the inclusion of the higher-order terms now become
dynamical.  In particular, we find that certain four-dimensional
vector supermultiplets
have wave operators that give rise to masses $m$ that are determined by a
non-trivial polynomial of fourth order in $m^2$.  This leads to mass-squared
values that are not simply linear in the eigenvalues of the Laplace operators
on the internal space, but, rather, involve non-trivial roots of the
associated quartic equation.  One consequence of this is that the values
of $m^2$ can be negative or even complex, thus implying that there will be
instabilities.

   The occurrence of such states might at first sight seem surprising in a
supersymmetric vacuum.  A standard argument for positive semi-definiteness
of the energy, first given in \cite{Deser:1977hu}, uses the fact that if a state
$|\psi\rangle$ is annihilated by the supercharge $Q$, then the superalgebra
$\{Q,Q\}\sim P$ implies that $P_0\ge0$.  However, a crucial ingredient in
this argument is that the norm on the states $|\psi\rangle$ is positive
definite \cite{Boulware:1985nn} In our case, the higher-derivative terms in the six-dimensional
theory lead to ghost modes in the spectrum, and thus the assumptions required
for the positivity result in \cite{Deser:1977hu} are violated.

  The detailed structure of the quartic polynomial in $m^2$ for the
vector multiplets implies that two of the four roots are always real and
positive, while the remaining two can be complex.  The conditions under
which this occurs are governed by the ratio $M^2/g^2$ and by the Kaluza-Klein
level number $\ell$ of the harmonics on $S^2$. As $M^2$ becomes larger, the
non-positivity and complexity of the two roots sets in at larger and larger
values of the level number $\ell$.  $M^2$ must at least satisfy
$M^2\ge 8(5+2 \sqrt6)g^2$ in order for the roots to be real and positive
even at the lowest level $\ell=0$.

   We also study the spectrum of the modes in the non-supersymmetric
Minkowski$_4\times S^2$ vacua that arise for $S^2$ monopole charges $q$
greater than 1.  An interesting feature in these cases is that the
spectrum includes an $SU(2)$ multiplet of massless spin-$\ft32$ fields
at level $\ell=\ft12(|q|-3)$.

The organisation of the paper is as follows.  In section 2 we review the
six-dimensional gauged ${\cal N}=(1,0)$ off-shell $R+ |\hbox{Riem}|^2$ supergravity
that was recently constructed in \cite{Coomans:2012ax}.  In section 3 we study the complete
linearised spectrum of Kaluza-Klein modes in the supersymmetric
Minkowski$_4\times S^2$ vacuum, which has a monopole charge $q=1$ on
the $S^2$ internal space, and exists for any value of the coupling
$M^{-2}$ of the Riemann-squared invariant.  In section 4 we repeat the
analysis for the non-supersymmetric Minkowski$_4\times S^2$ vacua, which
have arbitrary integer monopole charge $|q|\ge 2$, and which
exist only for a special value of the ratio $g^2/M^2$.  For this analysis
we need many results on the properties of spin-weighted
spherical harmonics on $S^2$, since these are needed for the expansions
in the monopole background of the fermion fields and certain vector fields
that carry charges. We present a detailed discussion of these
harmonics in appendix B.  In appendix A we give our spinor conventions, and
in appendix C we summarise some results for spin projection operators in
four dimensions.


\section{The Theory}


The off-shell $6D$ $(1,0)$ supergravity multiplet consists of the fields \cite{Bergshoeff:1985mz}
\be
\left(e_\mu{}^a,\ V_\mu'^{ij},\ V_\mu, \ B_{\mu\nu},\ L,\ C_{\mu\nu\rho\sigma},\ \psi_\mu^i,\ \chi^i \right)
\ee
where $V_\mu'^{ij}$ is symmetric and traceless in its $Sp(1)$ doublet indices, $B$ and $C$ are antisymmetric tensor fields, $L$ is a real scalar, and the spinors are symplectic Majorana-Weyl.
The above fields have $(15,12,5,10,1,5, 40,8)$ degrees of freedom. In addition, we shall consider the off-shell Maxwell multiplet consisting of the fields
\be
\left( A_\mu, \ Y^{ij},\ \lambda^i\right)\ ,
\ee
where $Y^{ij}$ is symmetric in its indices and the fermion is symplectic Majorana Weyl. These fields have $(5,3,8)$ degrees of freedom.

The total Lagrangian we shall study is given by
\be
{\cal L}= {\cal L}_R -\frac{1}{8M^2} {\cal L}_{R^2}\ ,
\ee
where the $U(1)_R$ gauged off-shell supergravity Lagrangian, up to quartic fermion terms, is
\cite{Bergshoeff:1985mz,Coomans:2012ax}\footnote{We have let $g\rightarrow 4g$ and $A_\mu \rightarrow A_\mu/{\sqrt 2}$
in the results of \cite{Coomans:2012ax}.}
\bea
e^{-1}{\cal L}_R &=&
\frac12 L R +\frac12 L^{-1}\partial_\mu L\partial^\mu L
+2{\sqrt{2}}g L \delta^{ij}Y_{ij} -\frac{1}{24} L H_{\mu\nu\rho}H^{\mu\nu\rho}
\nn\\
&&  +L  V'_\mu{}^{ij}V^{'\mu}{}_{ij} -\frac14 L^{-1}E^\mu E_\mu
+\frac{1}{\sqrt 2} E^\mu \left( V_\mu+2g A_\mu \right)
\nn\\
&& +Y^{ij} Y_{ij}-\frac{1}{8}F_{\mu\nu}F^{\mu\nu}
-\frac{1}{16}\varepsilon^{\mu\nu\rho\sigma\lambda\tau}B_{\mu\nu}F_{\rho\sigma}F_{\lambda\tau}
\nn\w2
&& -\frac12L \opsi_{\rho}\gamma^{\mu\nu\rho}D_{\mu}\psi_{\nu}
-{\sqrt 2} {\bar\chi}_i\gamma^{\mu\nu}D_\mu\psi_{\nu j} \delta^{ij}
+ L^{-1}{\bar\chi} {\slashed D}\chi
\nn\\
&& -\frac12 {\bar\psi}^\mu \gamma^\nu\psi_\nu  \partial_\mu L
-\frac{1}{\sqrt 2} \delta_{ij}{\bar\psi}_\nu^i\gamma^\mu\gamma^\nu\chi^j L^{-1}\partial_\mu L
-2{\sqrt 2}g L {\bar\lambda}_i\gamma^\mu\psi_{\mu j} \delta^{ij}
\nn\w2
&&
+ 2 g {\bar\lambda}\chi
+\frac12 V_\mu^{'ij}\left( {2\sqrt 2}{\bar\chi}^k\psi^\mu_i\delta_{jk}
-3L^{-1}{\bar\chi}_i\gamma^\mu\chi_j\right)
\nn\\
&&  -\frac{1}{48}L H_{\mu\nu\rho}
\left(\opsi^\lambda\gamma_{[\lambda} \gamma^{\mu\nu\rho}\gamma_{\tau]}\psi^\tau
+2{\sqrt 2} L^{-1}{\bar\psi}_{\lambda i}\gamma^{\lambda\mu\nu\rho}\chi_j \delta^{ij}
-2L^{-2}{\bar\chi}\gamma^{\mu\nu\rho}\chi\right)
\nn\\
&& -\frac{1}{4\sqrt 2} E_\rho
\left(\psi_\mu^i \gamma^{\rho\mu\nu}\psi_\nu^j \delta_{ij}-2{\sqrt 2} L^{-1} {\bar\psi}_\sigma \gamma^{\rho}\gamma^{\sigma}\chi
+2 L^{-2}{\bar\chi}_i\gamma^\rho\chi_j \delta^{ij} \right)
\nn\\
&&-2{\bar\lambda}{\slashed D}\lambda
+\frac{1}{12}H_{\mu\nu\rho}{\bar\lambda}\gamma^{\mu\nu\rho}\lambda
+\frac1{2\sqrt {2} } F_{\mu\nu} {\bar\lambda}\gamma^\rho \gamma^{\mu\nu}\psi_\rho\ ,
\label{Lag2}
\eea
where $H_{\mu\nu\rho}=3\partial_{[\mu} B_{\nu\rho]}$ and
\bea
E^\mu &=& \frac{1}{24}\varepsilon^{\mu\nu_1\cdots\nu_5} \partial_{[\nu_1}C_{\nu_2\cdots \nu_5]}\ .
\label{defH}\w2
D_\mu\psi_\nu^i &=& (\partial_\mu + \frac14\omega_{\mu}{}^{ab}\gamma_{ab} )\psi_\nu^i
-\frac12 V_\mu\delta^{ij}\psi_{\nu j}\ ,
\label{cd1}\w2
D_\mu\chi^i &=& (\partial_\mu + \frac14\omega_{\mu}{}^{ab}\gamma_{ab})\chi^i
-\frac12 V_\mu\delta^{ij}\chi_j + V_\mu{}^{'i}{}_j \chi^j\ .
\label{cd2}
\eea
Note the presence of arbitrary coupling constant in ${\cal L}_R$. In fact, the sum of all the terms in this Lagrangian that depend on $g$ separately have the off-shell supersymmetry. Thus, the total Lagrangian is a sum of three separately off-shell supersymmetric pieces.

The Lagrangian for the supersymmetrized Riemann squared term, up to quartic fermion terms, is given by
\cite{Bergshoeff:1986wc,Bergshoeff:2011xn}
\bea
e^{-1} {\mathcal L}_{\rm R^2} &=& R_{\mu\nu}{}^{ab}(\omega_-)R^{\mu\nu}{}_{ab}(\omega_-)
-2G^{ab}G_{ab} -4G_{\mu\nu}^{'ij}G^{'\mu\nu}_{ij}\ ,
\nonumber \\
&& +\frac14 \varepsilon^{\mu\nu\rho\sigma\lambda\tau}B_{\mu\nu}
R_{\rho\sigma ab}(\omega_-)R_{\lambda\tau}{}_{ab}(\omega_-)
\nn\w2
&& +2{\bar\psi}^{ab}(\omega_+)\gamma^\mu D_\mu (\omega,\omega_-)\psi_{ab}(\omega_+)
- R_{\nu\rho}{}^{ab}(\omega_-){\bar\psi}_{ab}(\omega_+)\gamma^\mu \gamma^{\nu\rho}\psi_\mu
\nn\w2
&& -8G_{\mu\nu}^{ij}\left({\bar\psi}^\mu_i \gamma_\lambda \psi^{\lambda\nu}_j(\omega_+)
+\frac16 {\bar\psi}^\mu_i \gamma\cdot H\psi^\nu_j\right)
 -\frac{1}{12}{\bar\psi}^{ab}(\omega_+)\gamma\cdot H \psi_{ab}(\omega_+)
\nn\\
&&-\frac12 \Bigl[ D_\mu(\omega_-,\Gamma_+) R^{\mu\nu ab} (\omega_-)
-2 H_{\mu\nu}{}^\rho R^{\mu\nu ab} (\omega_-)\Bigr] {\bar\psi}^a\gamma_\rho\psi_b\ ,
\label{R2}
\eea
where  $G_{\mu\nu}^{'ij}$ and $G_{\mu\nu}$ are the field strengths associated with $V_\mu^{'ij}$ and $V_\mu$, which can be combined as $V_\mu^{ij} = V_\mu^{'ij} + \ft12 \delta^{ij} V_\mu$. Furthermore $\psi_{\mu\nu}(\omega_+) =2D_{[\mu}(\omega_+) \psi_{\nu]}$ and
\bea
D_\mu (\omega,\omega_-)\psi^{ab i} &=&
(\partial_\mu +\frac14\omega_\mu{}^{cd}\gamma_{cd} )\psi^{ab i}
+2 \omega_{\mu -}{}^{c[a} \psi^{b]i}{}_c  +V_\mu^i{}_j \psi^{ab i}\ ,
\nn\w2
\omega_{\mu\pm}{}^{ab} &=& \omega_\mu{}^{ab} \pm \ft12 H_\mu{}^{ab}\ ,\qquad
\Gamma_{\mu\nu\pm}^\rho =\Gamma_{\mu\nu}^\rho \pm \ft12 H_\mu{}^{\nu\rho}\ .
\eea

The off-shell resulting supersymmetry transformations of the Poincar\'e multiplet, up to cubic fermion terms, are
\cite{Bergshoeff:1985mz,Bergshoeff:1986wc,Coomans:2012ax}
\bea
\delta e_{\mu}{}^a&=&\frac12{\bar\epsilon}\gamma^a\psi_{\mu}\ ,
\nn\\
\delta \psi_{\mu}{}^i&=& (\partial_{\mu} +\frac14\omega_{\mu ab}\gamma^{ab})\epsilon^i
+V_\mu{}^i{}_j\epsilon^j +\frac18 H_{\mu\nu\rho}\gamma^{\nu\rho}\epsilon^i\ ,
\nn\\
\delta B_{\mu\nu}&=&-{\bar\epsilon}\gamma_{[\mu}\psi_{\nu]}\ ,
\nn\\
\delta \chi^i &=& \frac{1}{2\sqrt 2} \gamma^\mu \delta^{ij}\partial_\mu L \epsilon_j
-\frac14 \gamma^\mu E_\mu\epsilon^i
+\frac{1}{\sqrt2}\gamma^\mu V'{}_\mu^{(i}{}_k \delta^{j)k} L
\epsilon_j - \frac{1}{12\sqrt2}L\delta^{ij}\gamma\cdot H \epsilon_j \ ,
\nn\w2
\delta L &=& \frac{1}{\sqrt 2} {\bar\epsilon}^i \chi^j\delta_{ij} \ ,
\nn\w2
\delta C_{\mu\nu\rho\sigma} &=& L{\bar\epsilon}^i\gamma_{[\mu\nu\rho}\psi_{\sigma]}^j\delta_{ij} -\frac{1}{2\sqrt 2} {\bar\epsilon}\gamma_{\mu\nu\rho\sigma}\chi \ ,
\nn\w2
\delta V_{\mu}{}^{ij}&=& \frac{1}{2}\bar{\epsilon}^{(i}\gamma^{\rho}\psi_{\mu\rho}^{j)}+\frac{1}{12}\bar{\epsilon}^{(i}\gamma\cdot H\psi_{\mu}^{j)} +\frac{1}{8}\sigma^{-1}\bar{\epsilon}^{(i}\gamma^{\rho}\Bigl(H_{[\mu}{}^{ab}\gamma_{ab}\psi_{\rho]}^{j)}\Bigr)
\label{susy2}
\eea
and the off-shell supersymmetry transformations of the vector multiplet are
\begin{eqnarray}
\delta A_{\mu}&=&-\bar{\epsilon}\gamma_{\mu}\lambda\ ,
\nonumber \\
\delta \lambda^{i}&=&\frac{1}{8\sqrt 2}\gamma^{\mu\nu}F_{\mu\nu}\epsilon^i-\frac{1}{2}Y^{ij}\epsilon_j\ ,
\nonumber \\
\delta Y^{ij} &=& -\bar{\epsilon}^{(i}\gamma^{\mu}D_{\mu}\lambda^{j)}
+\frac18 {\bar\epsilon}^{(i} \gamma^\mu\gamma\cdot H \psi_\mu^{j)}
-\frac{1}{24} {\bar\lambda}^i \gamma\cdot H \lambda^j
-\frac12 Y^{k(i} {\bar\epsilon}^{j)} \gamma^\mu \psi_{\mu k}\ .
\end{eqnarray}
Of the auxiliary fields of the Poincar\'e supergravity, $V_\mu^{'ij}$ and $V_\mu$ can no longer be eliminated algebraically due to the presence of the Riemann squared invariant but $Y^{ij}$ and $C_{\mu\nu\rho\sigma}$ can still be eliminated by means of their field equations as
\be
Y^{ij} = -{\sqrt 2} g L\delta^{ij}\ ,\qquad
E_\mu = {\sqrt 2} L\left(V_\mu+2g A_\mu\right)\ .
\ee
The total Lagrangian we shall study here is given by
\be
{\cal L}= L_{\rm R} -\frac1{8M^2} {\cal L}_{\rm R^2}\ ,
\ee
where $M$ is an arbitrary mass parameter.


\section{Spectrum in Supersymmetric $\rm{Minkowski}_4\times S^2$ Background }



\subsection{Supersymmetric $\rm{Minkowski}_4\times S^2$ background}


We shall study the compactification on the one half supersymmetric vacuum solution with the geometry of $\rm{Minkowski}_4\times S^2$. From here on, the 6D coordinates will be denoted by $x^M$ and they will be split as $(x^\mu,y^m)$ to denote the coordinates of 4D spacetime and the internal two-dimensional space. The supersymmetric $\rm{Minkowski}_4\times S^2$ vacuum solution given by \cite{Coomans:2012ax}
\begin{alignat}{3}
 \bar{R}_{\mu\nu\lambda\rho} &=0\ , &\qquad  \bar{R}_{mn} &= \alpha^2\bar{g}_{mn}\ , \qquad \bar{L} &=1\ , \nonumber\\
  \bar{F}_{\mu\nu} &=0\ , & \qquad  \bar{F}_{mn} &= 4g\epsilon_{mn}\ , \qquad &  \nonumber\\
  \bar{G}_{\mu\nu} &=0\ , & \qquad \bar{G}_{mn} &= -\alpha^2\epsilon_{mn}\ , \qquad &
\label{gb}
\end{alignat}
where $\alpha^2\equiv8g^2$, $\bar{g}_{mn}$ is the metric on $S^2$ with radius 1/$\alpha$, and $\epsilon_{mn}$ is the Levi-Civita tensor on the same $S^2$. We define the complex vectors
\be
{\hat Z}_M = {\hat V}^{'11}_M+i {\hat V}^{'12}_M\ ,
\ee
and parametrize the linearized fluctuations around above background as follows
\begin{alignat}{3}
   \hat{g}_{MN} &=\bar{g}_{MN}+\hat{h}_{MN}\ ,&\qquad \hat{L} &=1+\hat{\phi}\ ,&\qquad \hat{A}_{M} &=\bar{A}_M+\hat{a}_M\ ,\nonumber\\
   \hat{V}_{M} &=\bar{V}_M+\hat{v}_{M}\ , &\qquad ~~{\hat Z}_{M} &={\hat z}_M\ , &\qquad \hat{B}_{MN} &=\hat{b}_{MN}\ ,
\end{alignat}
where we use ``$\rm{hat}$'' to stand for six dimensional quantities and ``$\rm{bar}$'' to denote quantities evaluated in the vacuum background. In the background specified above, the linearized six dimensional bosonic and fermionic gauge symmetries are expressed as\footnote{For later convenience, starting from the USp(2) symplectic-Majorana-Weyl spinors we have defined Weyl spinors by complexifying as $\psi=\psi_1+i\psi_2 $ and rescaled $\hat{\chi}$ and $\hat{\lambda}$ used in \cite{Coomans:2012ax} by $\hat{\chi}\rightarrow\sqrt{2}\hat{\chi}$, $\hat{\lambda}\rightarrow\sqrt{2}\hat{\lambda}$. }
\begin{alignat}{3}
  \delta \hat{h}_{MN} &= \bar{\nabla}_{M}\hat{\xi}_N+ \bar{\nabla}_{N}\hat{\xi}_M\ , & \qquad \delta \hat{ a}_{M} &=\hat{\xi}^{N}\bar{F}_{NM}+\partial_{M}\hat{\Lambda}, & \nonumber\\
  \delta \hat{v}_{M}&=\hat{\xi}^{N}\bar{G}_{NM}-2g\partial_{M}\hat{\Lambda}\ , &\qquad \delta \hat{b}_{MN} &=\partial_{M}\hat{\Lambda}_{N}-\partial_{N}\hat{\Lambda}_{M}\ , & \nonumber\\
  \delta\hat{\psi}_M &=\bar{D}_M\hat{\epsilon}, &\qquad \delta\hat{\lambda} &=\ft{1}{16}{\bar\Gamma}^{MN}\bar{F}_{MN}\hat{\epsilon}+\ft{i}{2}g\hat{\epsilon}\ ,  & \qquad
  \delta\hat{\chi}& =0\ .
\label{6dgs}
\end{alignat}
This background preserves half supersymmetry because it admits a Killing spinor $\hat{\eta}$ which has the following properties
\be
\delta\hat{\psi}_M=\bar{D}_M\hat{\eta}=0,\qquad \delta\hat{\chi}=0,\qquad
\delta\hat{\lambda}=(\ft{1}{16}{\bar\Gamma}^{MN}\bar{F}_{MN}+\ft{i}{2}g)\hat{\eta}=0\ ,
\ee
and by choosing the six dimensional gamma matrices as in Appendix, it can be shown that
\be
\hat{\eta}=\epsilon\otimes\eta,\qquad \eta=\left(
                                             \begin{array}{c}
                                               0 \\
                                               1 \\
                                             \end{array}
                                           \right),
\ee
where $\epsilon$ is a constant four dimensional Weyl spinor with appropriate chirality inherited from six dimensions.


\subsection{Bosonic Sector}


In this section, we shall drop the ``bar'' on the covariant derivatives for simplicity in notation. The linearized bosonic field equations are given as follows\\
\begin{eqnarray}
 (\hat{R}^{(L)}_{MN}+\hat{\phi}\bar{R}_{MN}) &=&{\hat\nabla}_M{\hat\nabla}_N\hat{\phi}+\alpha^2\bar{g}_{MN}\hat{\phi}
  +\ft{1}{2}(\hat{F}^{(L)}_{MP}\bar{F}_{N}^{~P}+ \hat{F}^{(L)}_{NP}\bar{F}_{M}^{~P} -\bar{F}_{MP}\bar{F}_{NQ}\hat{h}^{PQ})
\w2
   &&+\ft{1}{2}\hat{h}_{MN}(\alpha^2-\ft{1}{4}\bar{F}^{PQ}\bar{F}_{PQ})
   -\ft{1}{4}\bar{g}_{MN}(\hat{F}^{(L)}_{PQ}\bar{F}^{PQ}-\bar{F}_{P}^{~Q}\bar{F}^{PT}\hat{h}_{QT}),\nn\\
   &&-\ft{1}{8M^2}S^{(L)}_{MN}\ ,
\nonumber\w2
\hat{R}^{(L)} &=& 2\alpha^2\hat{\phi}+2\hat{\Box}\hat{\phi}\ ,
\w2
\hat{\nabla}^{P}\hat{H}^{(L)}_{PMN} &=& \ft{1}{2}\varepsilon_{MN}^{~~~~PQST}(\ft{1}{2}\hat{F}^{(L)}_{PQ}\bar{F}_{ST}
 -\frac{1}{2M^2}\tilde{R}^{(L)J}_{~~~~KPQ}\bar{R}^{K}_{~JST})+\frac{1}{M^2}\hat{\nabla}^P\hat{\Box} \hat{H}^{(L)}_{PMN}, \nonumber\\
&&+3{M^2}\hat{\nabla}^P(\hat{H}^{(L)ST}_{+~~~~[P}\bar{R}_{MN]ST})\ ,
\end{eqnarray}
\bea
    0&=& \hat{\nabla}^{P}\hat{F}^{(L)}_{PM}-\hat{\nabla}^P\hat{h}_{PQ}\bar{F}^{Q}_{~M}
    -\hat{\nabla}_P\hat{h}_{QM}\bar{F}^{PQ}
    +\ft{1}{2}\hat{\nabla}_P\hat{h}\bar{F}^{P}_{~M}+4g(\hat{v}_M+2g\hat{a}_M)
    -\ft{1}{2}*\hat{H}^{(L)}_{MPQ}\bar{F}^{PQ}\ ,
\nn\\
&&\\
0&=& \hat{\nabla}^P\hat{G}^{(L)}_{PM}-\hat{\nabla}^P\hat{h}_{PQ}\bar{G}^{Q}_{~M}
-\hat{\nabla}_P\hat{h}_{QM}\bar{G}^{PQ}
    +\ft{1}{2}\hat{\nabla}_P\hat{h}\bar{G}^{P}_{~M}-M^2({\hat v}_M+2g {\hat a}_M) \ ,
\w2
0&=& \left( {\hat\nabla}^P - i{\bar V}^P\right) \hat{G}_{PM}^{'(L)}- i{\bar G}_{MN} {\hat z}^N-M^2 {\hat z}_M \ ,
\eea
where
\bea
{\hat G}_{MN}^{'(L)} &=& 2{\hat D}_{[M}{\hat z}_{N]}\ ,\qquad
{\hat D}_M {\hat z}_N \equiv ( {\hat\nabla}_M - i{\bar V}_M ){\hat z}_N\ ,
\nn\w2
R^{(L)P}_{~~~~MNQ} &=&
   \hat{R}^{(L)P}_{~~~~MNQ}-\hat{\nabla}_{[N}H^{(L)P}_{Q]~~M}\ ,
\nn\w2
S^{(L)}_{MN} &=& 8\left(\hat{G}^{(L)}_{MP}\bar{G}_{N}^{~P}+\hat{G}^{(L)}_{NP}\bar{G}_{M}^{~P}
    -\bar{G}_{M}^{~P}\bar{G}_{N}^{~Q}\hat{h}_{PQ}\right)
    -4\bar{g}_{MN}(\hat{G}^{(L)}_{PQ}\bar{G}^{PQ}-\bar{G}_{P}^{~Q}\bar{G}^{PT}\hat{h}_{QT})
\nonumber\\
    && +4(\tilde{R}^{(L)S}_{~~~~QMP}\bar{R}^{Q~~~P}_{~SN}+\bar{R}^{S~~~P}_{~QM}\tilde{R}^{(L)Q}_{~~~~SNP}
    -\hat{h}^{PQ}\bar{R}^{S}_{~TMP}\bar{R}^{T}_{~SNQ})-2\hat{h}_{MN}\bar{G}^{PQ}\bar{G}_{PQ} \nonumber\\
    &&+\hat{h}_{MN}\bar{R}^{PQST}\bar{R}_{PQST}-2\bar{g}_{MN}(\tilde{R}^{(L)S}_{~~~~TPQ}\bar{R}^{T~PQ}_{~S} +\bar{R}_{JKSP}\bar{R}^{JKS}_{~~~~Q}\hat{h}^{PQ})
\nonumber\\
    &&+8(\hat{\nabla}^{P}\tilde{\nabla}^{Q}\tilde{R}_{P(MN)Q})^{(L)}
    +8\hat{\nabla}^{S}(\bar{R}_{S(M}^{~~~~PQ}\hat{H}^{+{(L)}}_{N)PQ}),
\end{eqnarray}
and the penultimate term takes the form
\bea
(\hat{\nabla}^{P}\tilde{\nabla}^{Q}\tilde{R}_{P(MN)Q})^{(L)} &=&\bar{R}^{P~~~Q}_{~(MN)}(\bar{R}_{P}^{~S}\hat{h}_{SQ}-\bar{
  R}^{S~T}_{~P~Q}\hat{h}_{ST}-\ft{1}{2}\hat{\Box} \hat{h}_{PQ})+\hat{\nabla}_{P}\hat{\nabla}_{Q}\hat{h}_{S(M}\bar{R}^{P~~QS}_{~N)}\nn\\
     &&-\ft{1}{2}(\hat{\nabla}^{P}\hat{\nabla}_{(M}\hat{h}^{QS}\bar{R}_{N)SPQ}+
     \hat{\nabla}^{P}\hat{\nabla}^{S}\hat{h}^{Q}_{~(M}\bar{R}_{N)SPQ}\nn\\
&&-\ft{1}{2}\bar{R}^{PQS}_{~~~~~T}\hat{h}^{T}_{~(M}\bar{R}_{N)SPQ}
+\ft{1}{2}\bar{R}^{PQ}_{~~~T(M}\bar{R}_{N)SPQ}\hat{h}^{ST} )\nn\\
   &&  +\ft{1}{2}\bar{R}_{PMNQ}\hat{\nabla}^{P}\hat{\nabla}^{Q}\hat{h}+\hat{\nabla}_{P}\hat{\nabla}^{Q}\tilde{
   R}^{(L)P}_{~~~~(MN)Q}\nn\\
   && -\ft{1}{2}(\hat{\nabla}_P \hat{H}^{(L)}_{QS(M}\bar{R}^{PQ~S}_{~~~N)}+\hat{\nabla}_P\hat{H}^{(L)}_{QS(M}\bar{R}^{P~QS}_{~N)}).
\eea
The covariant derivative ${\tilde\nabla}_M$ is defined with respect to the connection ${\widetilde\Gamma}_{\mu\nu}^\rho$ containing bosonic torsion as
\be
{\widetilde\Gamma}_{\mu\nu}^\rho = \left\{
                        \begin{array}{c}
                          \rho \\
                          \mu\nu \\
                        \end{array}
                      \right\} +\frac12 H_{\mu\nu}{}^\rho\ .
\ee
Note that we are using $\hat G^{'(L)}_{MN}$ to denote the covariant field
strength of the complex vector field $\hat z_M$, and $G^{(L)}_{MN}$
to denote the field strength of the real vector $v_M$.

There are no transverse traceless spin-2 harmonics on $S^2$,
and the transverse spin-1 harmonics are related to spin-0 harmonics by
\be
Y^{(\ell)}_m=\epsilon_m{}^n\nabla_nY^{(\ell)}\ .
\label{ym}
\ee
We can expand the six-dimensional bosonic fields in terms of $S^2$
harmonics as follows
\begin{eqnarray}
   \hat{h}_{\mu\nu}& =&\sum_{\ell\geq0 }h_{\mu\nu}^{(\ell)}Y^{(\ell)} \ ,
\nn\\
   \hat{h}_{mn}&=&\sum_{\ell\geq2 }\left(L^{(\ell )}\nabla_{\{m}Y^{(\ell)}_{n\}}+\tilde{L}^{(\ell)}\nabla_{\{m}\nabla_{n\}}Y^{(\ell)}\right)
   + \bar{g}_{mn}\sum_{\ell \geq0}N^{(\ell)}Y^{(\ell)}\ ,
\nn\\
  \hat{h}_{\mu m}&=&\sum_{\ell\geq1}(k_{\mu}^{(\ell)}Y_{m}^{(\ell)}+\tilde{k}_{\mu}^{(\ell)}\nabla_mY^{(\ell)})\ ,
\nn\\
\hat{\phi} &=&\sum_{\ell\geq0}\phi^{(\ell)}Y^{(\ell)}\ ,
\nn\\
  \hat{a}_{\mu}&=&\sum_{\ell \geq0 }a_{\mu}^{(\ell)}Y^{(\ell)},\qquad \hat{a}_{m}=\sum_{\ell\geq1}\left(a^{(\ell)}Y_m^{(\ell)}+\tilde{a}^{(\ell)}\nabla_mY^{(\ell)}\right)\ ,
\nn\\
    \hat{v}_{\mu}&=&\sum_{\ell \geq0}v^{(\ell)}_{\mu}Y^{(\ell)},\qquad \hat{v}_{m}=\sum_{\ell \geq1}\left(v^{(\ell)}Y^{(\ell)}_{m}+{\tilde v}^{(\ell)}\nabla_m Y^{(\ell)}\right)\ ,
\nn\\
{\hat z}_\mu &=& \sum_{\ell\ge 1} z^{(\ell)}_\mu {}_{-1}{Y}^{(\ell)} \ ,
\nn\\
{\hat z}_m =&& \sum_{\ell=0,1} z^{(\ell)} {}_{-1}V_{m}^{(\ell)}
+\sum_{\ell\ge 2} \left(z^{(\ell)} D_m {}_{-1}{ Y}^{(\ell)} +i{\tilde z}^{(\ell)} \epsilon_m{}^n D_n {}_{-1}{Y}^{(\ell)}\right)\ ,
\nn\\
\hat{b}_{\mu\nu}&=&\sum_{\ell \geq0}b_{\mu\nu}^{(\ell)}Y^{(\ell)},\qquad \hat{b}_{mn}=\epsilon_{mn}\sum_{\ell \geq0}b^{(\ell)}Y^{(\ell)}\ ,
\nn\\
  \hat{b}_{\mu m}&=&\sum_{\ell\geq1}\left(b_{\mu}^{(\ell)}Y^{(\ell)}_{m}
  +\tilde{b}_{\mu}^{(\ell)}\nabla_mY^{(\ell)}\right)\ ,
\label{he1}
\end{eqnarray}
where the notation $\{mn\}$ means ``symmetric and traceless,''
and in the $\hat z_m$ expansion
${}_{-1}V_{m}^{(0)}$ and ${}_{-1}V_{m}^{(1)}$ are level $\ell=0$
and $\ell=1$
complex anti-self dual vector harmonics with charge $-1$ on
the 2-sphere, whose explicit forms are given in Appendix B.2.
%
%
$D_m$ is the $U(1)$ covariant derivative on the 2-sphere, and
${}_{-1}{Y}^{(\ell)}$ are the charged harmonics which are described in
some detail in Appendix B.1. Furthermore, the scalar harmonics $Y^{(\ell)}$
employed above satisfy
\be
\label{}
    \Box_2Y^{(\ell)}=- \alpha^2\cell Y^{(\ell)}\ ,
\ee
where $\Box_2$ is the d'Alembertian on $S^2$ with radius $1/\alpha$ and
\be
\boxed{{\cell \equiv \ell(\ell+1)}} \ ,\qquad \boxed{\alpha^2\equiv 8g^2}\ .
\ee
We have also used the spin-1 harmonics $Y_m^{(\ell)}$ which satisfy the relations
\be
    \Box_2Y^{(\ell)}_n=-(c_\ell-1)\alpha^2 Y^{(\ell)}_n\ ,
    \qquad  \epsilon^{mn}\nabla_m Y^{(\ell)}_n=\alpha^2\cell\, Y^{(\ell)}\ .
\ee
Utilizing the six dimensional gauge symmetries (\ref{6dgs}), we impose the following gauge condition on the linearized fields
\cite{Kim:1985ez}
\bea\label{dedonder}
 \hat{\nabla}^{m}\hat{h}_{\{mn\}}&=&0,\qquad \hat{\nabla}^{m}\hat{h}_{m\mu}=0,\nn\\
  \hat{ \nabla}^m\hat{a}_m&=&0,\qquad \hat{ \nabla}^m\hat{b}_{mM}=0.
\eea
Upon the use of these gauge conditions, the harmonic expansions (\ref{he1}) simplify to
\begin{eqnarray}\label{HE}
    \hat{h}_{\mu\nu}&=&\sum_{\ell \geq0}h_{\mu\nu}^{(\ell)}Y^{(\ell)}\ ,\qquad\ \ \hat{h}_{\mu m}=\sum_{\ell\geq1}k_{\mu}^{(\ell)}Y_{m}^{(\ell)}\ ,
\nn\\
    \hat{h}_{mn}&=&\bar{g}_{mn}\sum_{\ell \geq0}N^{(\ell)}Y^{(\ell)},\qquad\hat{\phi}=\sum_{\ell \geq0}\phi^{(\ell)}Y^{(\ell)}\ ,
\nn\\
   \hat{a}_{\mu}&=&\sum_{\ell \geq0}a_{\mu}^{(\ell)}Y^{(\ell)}\ ,\qquad\quad  \hat{a}_{m}=\sum_{\ell \geq1}a^{(\ell)}Y^{(\ell)}_m\ ,
\nn\\
  \hat{v}_{\mu} & =&\sum_{\ell\geq0}v^{(\ell)}_{\mu}Y^{(\ell)}\ ,\qquad\quad \hat{v}_{m}=\sum_{\ell \geq1}\left(v^{(\ell)}Y^{(\ell)}_{m}+{\tilde v}^{(\ell)}\nabla_m Y^{(\ell)}\right)\ ,
\nn\\
{\hat z}_\mu &=& \sum_{\ell\ge 1} z^{(\ell)}_\mu\ _{-1} Y^{(\ell)} \ ,
\nn\\
{\hat z}_m =&& \sum_{\ell=0,1} z^{(\ell)} {}_{-1}V_{m}^{(\ell)}
+\sum_{\ell\ge 2} \left(z^{(\ell)} D_m {}_{-1}{Y}^{(\ell)} +i{\tilde z}^{(\ell)} \epsilon_m{}^n D_n{}_{-1}{Y}^{(\ell)}\right)\ ,
\nn\\
\hat{b}_{\mu\nu}&=&\sum_{\ell \geq0}b_{\mu\nu}^{(\ell)}Y^{(\ell)}\ ,\qquad\ \  \hat{ b}_{\mu m}=\sum_{\ell \geq1}b_{\mu}^{(\ell)}Y^{(\ell)}_{m}\ ,\qquad \hat{b}_{mn}=\epsilon_{mn}b^{(0)}Y^{(0)}\ .
\end{eqnarray}
The de Donder-Lorentz gauge (\ref{dedonder}) does not fix all the gauge symmetries, and consequently there are some residual ones generated by harmonic zero modes, $S^2$ Killing vector $Y^{(1)}_m$ and conformal Killing vectors $\nabla_mY^{(1)}$. Specifically, these residual gauge symmetries are:
\begin{itemize}

\item   The four dimensional coordinate transformation generated by $\hat{\xi}_{\mu}=\xi^{(0)}_{\mu}Y^{(0)}$
         \begin{equation}\label{bs-1}
           \delta h_{\mu\nu}^{(0)}=\partial_{\mu}\xi^{(0)}_{\nu}+\partial_{\nu}\xi^{(0)}_{\mu}\ .
         \end{equation}

\item  The Stueckelberg shift symmetries generated by $\hat{\xi}_{m}=\xi^{(1)}\nabla_{m}Y^{(1)}$
        \begin{eqnarray}\label{bs-2}
          \delta h^{(1)}_{\mu\nu}&=&-\partial_{\mu}\partial_{\nu}\xi^{(1)},\qquad \delta N^{(1)}=-2\xi^{(1)} \nn \\
          \delta a^{(1)}&=&4g\xi^{(1)},\qquad \delta v^{(1)}=-\alpha^2\xi^{(1)}\ .
        \end{eqnarray}

\item    Linearized $SU(2)$ symmetry generated by $\hat{\xi}_{m}=\xi'^{(1)}Y^{(1)}_m$
             and $\hat{\Lambda}=-4g\xi'^{(1)}Y^{(1)}$

             \begin{equation}\label{bs-3}
            \delta k_{\mu}^{(1)}=\partial_{\mu}\xi'^{(1)}\ .
             \end{equation}

\item    Four dimensional $U(1)_{\rm{R}}$ symmetry generated by $\hat{\Lambda}=\Lambda^{(0)} Y^{(0)}$

        \begin{equation}\label{bs-4}
            \delta a^{(0)}_{\mu}=\partial_{\mu}\Lambda^{(0)}\ .
        \end{equation}

\item  Abelian 2-form symmetry generated by $\hat{\Lambda}_{\mu}=\Lambda^{(0)}_{\mu} Y^{(0)}$

         \begin{equation}\label{bs-5}
            \delta b_{\mu\nu}^{(0)}=\partial_{\mu}\Lambda_{\nu}^{(0)}-\partial_{\nu}\Lambda_{\mu}^{(0)}\ .
        \end{equation}

\end{itemize}
We shall take into account these symmetries in the analysis of the spectrum below, where we treat the spin-2, spin-1 and spin-0 sectors separately. In doing so we shall encounter the following wave operators
\bea
{\cal O}_1 &\equiv&\hat\Box_0 + \alpha^2 - M^2\ ,
\nn\w2
{\cal O}_2  &\equiv&  \hat\Box_0^2 -M^2 \hat\Box_0 -\alpha^4\,\cell \ ,
\nn\w2
{\cal O}_4 &\equiv& \hat\Box_0^4 + (\alpha^2-M^2)\, \hat\Box_0^3
-2\alpha^2(\alpha^2\cell - M^2)\, \hat\Box_0^2 -4\cell\,\alpha^4 (\alpha^2 - M^2)\, \hat\Box_0 -
2\alpha^8\, \cell^2\ ,
\label{1234}
\eea
where
\be
\boxed{\hat{\Box}_0 \equiv  \Box- \alpha^2\cell}\ .\label{boxhat}
\ee
In particular, the operator ${\cal O}_4$ has the property that for $\ell=1$ it factorizes as

\bea
{\cal O}_4|_{\ell=1}  &=& \Box\,{\cal O}_3\ ,
\nn\w2
{\cal O}_3 & \equiv& \Box^3 -(M^2+7\alpha^2)\Box^2 +2\alpha^2 (4M^2+7\alpha^2)\Box -12\alpha^4(M^2+\alpha^2)\ .
\label{op3}
\eea
In the $\ell=0$ sector, we will encounter the wave operator
\be
\widetilde{\cal O}_2 = \Box(\Box+\alpha^2) -M^2(\Box-2\alpha^2)\ .
\label{cal2}
\ee

\bigskip


\subsection*{\fbox{Spin-2 sector}}


\bigskip

The spin-2 sector contains only the transverse and traceless gravitons, which upon the use of the spin projector operators provided in the Appendix C, and for $\ell\ge 1$, satisfy the following equation
\be
\ell\ge 1:\qquad {\cal O}_2 \left(\cP^{2}h\right)^{(\ell)}_{\mu\nu}=0\ ,
\ee
where $P^2$ is the spin-2 projector defined in Appendix C. This equation describes two massive
gravitons with mass squared
\be
\ell\ge 1: \qquad m^2_{\pm} (\ell)= \frac12 \left(M^2 +2\alpha^2c_\ell \pm \sqrt{M^4+4\alpha^4 c_\ell}\right)\ .
\label{mass2}
\ee
The $\ell=0$ needs to be treated separately, and in this case the gravitons satisfy
\be
(\Box-M^2)R^{L(0)}_{\mu\nu}
=-M^2\left( \partial_{\mu}\partial_{\nu}S^{(0)}+\alpha^2\eta_{\mu\nu} S^{(0)}\right)
+\partial_{\mu}\partial_{\nu}(\Box+\alpha^2) S^{(0)},
\ee
where $S^{(0)}=\phi^{(0)}+N^{(0)}$. The solutions of this equation can be expressed as $h^{(0)}_{\mu\nu}=h'^{(0)}_{\mu\nu}+h''^{(0)}_{\mu\nu}$, where  $h''^{(0)}_{\mu\nu}$ is completely determined by $S^{(0)}$ while $h'^{(0)}_{\mu\nu}$ is the solution to the following equations modulo the gauge symmetry (\ref{bs-1}):
\be
\ell=0:\qquad \Box(\Box-M^2)h'^{(0)}_{\mu\nu}=0,\qquad\,R'^{(0)}=0\ ,
\ee
which describe a massless graviton and massive graviton with $m^2=M^2$.

\bigskip


\subsection*{\fbox{Spin-1 sector}}


\bigskip

Let $\ell\ge 2$. Then, the spin-1 sector consists of eight vectors $(P^{1}h)_{\mu\nu},\partial^{\nu}b_{\mu\nu}, z^{T}_{\mu}, k^{T}_{\mu},a^{T}_{\mu},v^{T}_{\mu},b_{\mu\nu}^{T},b^{T}_{\mu})$, where ``T'' indicates the transverse part and $(P^{1}h)_{\mu\nu}=P^1_{\mu\nu}{}^{\rho\sigma}h_{\rho\sigma}$ (see Appendix B). Of these eight vectors, $(k^{T}_{\mu},a^{T}_{\mu},v^{T}_{\mu},b_{\mu\nu}^{T},b^{T}_{\mu})$ have mixing with each other through the following equations
\bea
0&=&{\cal O}_2 b^{T(\ell )}_{\mu\nu}
    +4gM^2\star F^{(\ell)}_{\mu\nu}(a)-\alpha^4\cell\left(\star F^{(\ell)}_{\mu\nu}(k)-\star F^{(\ell)}_{\mu\nu}(b)\right)\ ,
\label{spin1-1}\\
0&=& {\cal O}_1 \hat{\Box}_0 b^{T(\ell )}_{\mu}
    +\ft{1}{2}\alpha^2\epsilon_{\mu}^{~\nu\lambda\rho}\partial_{\nu}b_{\lambda\rho}^{T(\ell )}\ ,
\label{spin1-2}\\
0&=&(\hat{\Box}_0+\alpha^2)a^{T(\ell )}_{\mu}
-4g\alpha^2\cell k^{T(\ell )}_{\mu}+4gv_{\mu}^{T(\ell )}-2g\epsilon_{\mu}^{~\nu\lambda\rho}\partial_{\nu}b_{\lambda\rho}^{T(\ell )}\ ,
\label{spin1-3}
\\
0&=&(\hat{\Box}_0-M^2)v^{T(\ell )}_{\mu}+\alpha^2\cell k^{T(\ell )}_{\mu}+\ft{1}{4}gM^2a^{(\ell)}_{\mu}\ ,
\label{spin1-4}\\
0&=&\biggl( {\cal O}_2+\alpha^2(\hat{\Box}_0-\alpha^2\cell)\biggr)k^{T(\ell )}_{\mu}+4gM^2a^{T(\ell )}_{\mu}+\ft12 \alpha^2\left(4 v^{T(\ell )}_{\mu}
-\epsilon_{\mu}^{~\nu\lambda\rho}\partial_{\nu}b_{\lambda\rho}^{T(\ell )}\right)\ .
\label{spin1-5}
\eea
Diagonalising the associated $5\times 5$ operator-valued matrix, we find that the modes are annihilated
by ${\cal O}_1^2\,{\cal O}_2\,{\cal O}_4$. In particular, the linear combinations with coefficients $(-1,-4g\alpha^2\cell/M^2,0,0,1)$ and $(2,4g\alpha^2\cell/M^2,1,0,0)$
are annihilated by ${\cal O}_1$. The remaining vectors, namely,
$ \left( (P^{1} h)_{\mu\nu},\partial^{\nu}b_{\mu\nu}, z_\mu \right)$ are separately annihilated by ${\cal O}_1$ as well. In summary, for $\ell\ge 2$ the total wave operator can be denoted by
\be
\ell \ge 2:\qquad {\cal O}^{(1)} = {\cal O}_1^6\,{\cal O}_2\,{\cal O}_4\ ,
\ee
implying six massive vectors with mass squared
\be
\ell \ge 2:\qquad m^2(\ell) = M^2+ \alpha^2 (c_\ell-1)\ ,
\label{mass1}
\ee
two massive vectors with mass squared defined in Eq.(\ref{mass2}) and four massive vectors whose squared masses are given by the roots of the polynomial
\bea
&& x^4 + ax^3+bx^2+cx+d=0\ ,
\nn\w2
&& a= -M^2-(4\ell^2+4\ell-1)\alpha^2\ ,
\nn\w2
&& b= \alpha^2\left[ 2M^2+\ell(\ell+1)\left(\,(6\ell^2+6\ell-5)\alpha^2+3M^2\right)\,\right] \ ,
\nn\w2
&& c=-\ell(\ell+1)\alpha^2 \left\{ \alpha^2\left[4+\ell(\ell+1)(4\ell^2+4\ell-7)\right]\alpha^2
+3\ell(\ell+1)M^2\right\}\ ,
\nn\w2
&& d=\ell^2(\ell+1)^2(\ell-1)(\ell+2)\alpha^6 \left[(\ell^2+\ell-1)\alpha^2+M^2\right]\ .
\label{mass4}
\eea

Next, consider the case $\ell=1$. Recalling the factorization result given in \eq{op3}, the total wave operator
becomes
\be
\ell=1:\qquad {\cal O}^{(1)} = {\cal O}_1^6 {\cal O}_2|_{\ell=1} \Box {\cal O}_3\ .
\label{p3}
\ee
In particular the  massless vector is a linear combination of $(k^{T(1)}_{\mu},a^{T(1)}_{\mu},v^{T(1)}_{\mu},b_{\mu\nu}^{T(1)},b^{T(1)}_{\mu})$ with mixing coefficients
$(1,-4g,\alpha^2,1,0)$. The squared masses associated with ${\cal O}_1^6 {\cal O}_2|_{\ell=1}$ can be read of from \eq{mass1} and \eq{mass2} by setting $\ell=1$, and those associated with ${\cal O}_3$ are the roots of the following polynomial
\be
x^3 -(M^2+7\alpha^2)x^2 +2\alpha^2 (4M^2+7\alpha^2)x -12\alpha^4(M^2+\alpha^2)=0\ .
\label{mass3}
\ee

There remains the case of $\ell=0$. The only vector fluctuations at this level are $(b_{\mu\nu}^{T(0)},a^{T(0)}_{\mu},v^{T(0)}_{\mu})$. Upon diagonalising the associated $3\times 3$ operator-valued matrix, we find that the modes are annihilated by the following partially-factorising operator polynomial
\be
\ell=0:\qquad {\cal O}^{(1)} = \Box(\Box-M^2) \widetilde{\cal O}_2 \ ,
\ee
where the would-be massless vector annihilated by $\Box$ is eaten by the two form and the operator ${\tilde O}_2$ is defined in \eq{cal2}. Thus, for $\ell=0$ there are no massless vector modes, a massive vector with mass $M$ and two massive vectors with squared masses given by
\be
\widetilde{m}^2_{\pm} = \frac12\left(M^2-\alpha^2 \pm \sqrt{M^4-10M^2\alpha^2 +\alpha^4}\, \right)\ .
\label{massT2}
\ee

\bigskip


\subsection*{\fbox{The Spin-0 sector}}


\bigskip

We start with the case $\ell\ge 2$. Defining $\tilde{\varphi}=\omega_{\mu\nu}h^{\mu\nu}$ and $\varphi=\ft13\theta_{\mu\nu}h^{\mu\nu}$ (see Appendix C), this sector consists of thirteen scalars $\left(\phi, N,\varphi, \tilde{\varphi},a, v, \partial^{\mu}k_{\mu},\partial^\mu b_{\mu},\partial^\mu a_{\mu},\partial^\mu v_{\mu}, \partial^\mu z_{\mu}, z, \tilde{z}\right)$.  The first six scalars
$(\phi,N, \varphi, \tilde{\varphi},a, v)$ mix as follows
\bea
\ell\ge 2:\ \ 0&=&2(\hat{\Box}_0+\alpha^2)\phi^{(\ell)}+(2\hat{\Box}_0+2\alpha^2+ \alpha^2\cell)N^{(\ell)}+3\hat{\Box}_0\varphi^{(\ell)}
    -\alpha^2\cell \tilde{\varphi}^{(\ell)}\ ,
\label{spin0-1}\\
0&=&{\cal O}_2\varphi^{(\ell)}+2\alpha^2M^2 \left[\phi^{(\ell)}-2g\cell a^{(\ell)}\right]-2\alpha^2(M^2-\alpha^2\cell)N^{(\ell)}+2\alpha^3\cell v^{(\ell)}\ ,
\label{spin0-2}\\
0&=&\biggl(M^2-(\alpha^2+\Box)\biggr)\tilde{\varphi}^{(\ell)} +3(M^2-\alpha^2)\varphi^{(\ell)}
+2M^2\phi^{(\ell)}+(2\hat{\Box}_0 +2\alpha^2+\alpha^2\cell)N^{(\ell)}\ ,
\nn\\
&& \label{spin0-3}\\
0&=& (\hat{\Box_0}+\alpha^2)a^{(\ell)}+4gN^{(\ell)}-2g(\tilde{\varphi}^{(\ell)}
+3\varphi^{(\ell)})+4gv^{(\ell)}\ ,
\label{spin0-4}\\
0&=&(\hat{\Box}_0-M^2)v^{(\ell)}-\alpha^2 N^{(\ell)}+\frac12\alpha^2(\tilde{\varphi}^{(\ell)} +3\varphi^{(\ell)})
-2 gM^2a^{(\ell)}\ ,
\label{spin0-5}\\
0&=&\biggl(M^2-(\alpha^2-\Box)\biggr)N^{(\ell)}
+3M^2\varphi^{(\ell)}+2M^2\phi^{(\ell)}-4gM^2a^{(\ell)}
-2\alpha^2 v^{(\ell)}-\alpha^2(\cell-1)\tilde{\varphi}^{(\ell)}\ .
\nn\\&&
\label{spin0-6}
\eea
Diagonalising the associated $6\times 6$ operator-valued matrix, we
find that the modes are annihilated by ${\cal O}_1^3\,{\cal O}_4$.
Of the remaining scalars, $(\partial^{\mu}k_{\mu},\partial^\mu b_{\mu},\partial^\mu a_{\mu},\partial^\mu v_{\mu},\tilde v)$ mix but only three of them are dynamical. We choose these to be $(\partial^{\mu}k_{\mu},\partial^\mu b_{\mu},\tilde v)$ which are separately annihilated by ${\cal O}_1$, while $(\partial^\mu a_{\mu},\partial^\mu v_{\mu})$ are determined by
\bea
\partial^\mu a_\mu^{(\ell)} &=& \frac{\alpha^2}{2g}\left( {\tilde v}^{(\ell)} - \partial^\mu k_\mu^{(\ell)}\right)\ ,
\nn\w2
\partial^\mu v_\mu^{(\ell)} &=& \alpha^2 (\cell-1){\tilde v}^{(\ell)} +\alpha^2 \partial^\mu k_\mu^{(\ell)}\ .
\eea
Finally, the remaining scalars $(z, \tilde{z})$ are annihilated by
${\cal O}_1$, and the longitudinal modes $\partial^{\mu}z_{\mu}$ are given
in terms of  $z$ and $\tilde{z}$, by virtue of the equation
$\hat{D}^{M}\hat{z}_M=0$. Thus, for $\ell\ge 2$ the total wave operator
is given by
\be
\ell \ge 2:\qquad {\cal O}^{(0)} = {\cal O}_1^{10}\,{\cal O}_4\ .
\ee
Of these, the three linear combinations of $(\phi, N,\varphi, \tilde{\varphi},a, v)$ with coefficients $(2+\alpha^2{\cell}/{M^2},-2,0,0,0,2)$, $(-2-\alpha^2\cell/{M^2},2,0,0,-{8g}/{M^2},0)$ and $(-2+\alpha^2{\cell}/{M^2},2,0,4,0,0)$ are annihilated by ${\cal O}_1$.

 In the case of $\ell=1$, utilizing the residual symmetry (\ref{bs-2}) and (\ref{bs-3}), one can eliminate $N^{(1)}$ and $\partial^\mu k_{\mu}^{(1)}$. Taking into account the fact that the harmonic expansion of ${\hat z}_m$ contributes only one complex scalar for $\ell=1$, namely $z^{(1)}$, we find that for $\ell=1$ the total wave operator for the scalar fields is given by
\be
\ell=1:\qquad {\cal O}^{(0)} = {\cal O}_1^6|_{\ell=1}\,{\cal O}_3\ .
\ee

There remains the case of $\ell=0$.   Of the remaining scalars, $(\phi^{(0)},N^{(0)},b^{(0)})$ satisfy the equations
$  \widetilde{\cal O}_2\,S^{(0)}=0$ where $S^{(0)}=\phi^{(0)}+N^{(0)}$, $\Box\,{{\cal O}_1}|_{\ell=0}\, b^{(0)}=0$ and $\Box\,{{\cal O}_1}|_{\ell=0}\, N^{(0)}= 0$. Finally, there is a complex scalar $z^{(0)}$ annihilated by ${\cal O}_1|_{\ell=0}$. Thus, for $\ell=0$ the total wave operator or the scalar fields is given by
\be
\ell=0: \qquad {\cal O}^{(0)} = \Box^2\,{\cal O}_1^4|_{\ell=0}\, \widetilde{\cal O}_2\ ,
\ee
implying two massless and six massive scalars in this sector.


\subsection{Fermionic sector}


In terms of the complex spinor, the linearized equations of fermions around the background (\ref{gb}) are given by
\bea
 0&=& {\bar\Gamma}^{MPQ}\bar{D}_{P}\psi_{Q}+2 i{\bar\Gamma}^{NM}\bar{D}_N\chi
-4g i{\bar\Gamma}^M\lambda
-\ft{1}{2}\bar{F}_{PQ}{\bar\Gamma}^{PQ}{\bar\Gamma}^{M}\lambda-\frac{1}{8M^2}\Theta^{M}\ ,
\\
 0&=&{\bar\Gamma}^{MN}\bar{D}_M\psi_{N}
-2i{\bar\Gamma}^{M}\bar{D}_{M}\chi-8gi\lambda,
\\
0&=&  2gi{\bar\Gamma}^M\psi_{M}+8g\chi-4{\bar\Gamma}^M {\bar D}_M\lambda
+\ft{1}{4}\bar{F}_{PQ}{\bar\Gamma}^{M}{\bar\Gamma}^{PQ}\psi_{M}\ ,
\eea
where
\bea
 \bar{D}_M\psi &=&(\partial_M+\ft{1}{4}\bar{\omega}_M^{~AB}{\bar\Gamma}_{AB})\psi-\ft{i}{2}\bar{V}_M\psi\ ,
\nn\w2
\Theta^{M}&=& 8{\bar\Gamma}^P\bar{D}_Q\bar{ D}_P\psi^{QM} -2\bar{R}^{PM}_{~~~ST}{\bar\Gamma}^Q{\bar\Gamma}^{ST}\bar{D}_{P}\psi_{Q}
+2\bar{R}^{PQ}_{~~~ST}{\bar\Gamma}^{ST}{\bar\Gamma}^{M}\bar{D}_{P}\psi_{Q}
\nn\\
&& +8i\bar{G}^{P[M}{\bar\Gamma}^{Q]}\bar{D}_{Q}\psi_{P}
-8i\bar{G}^{M[P}{\bar\Gamma}^{Q]}\bar{D}_{Q}\psi_{P}\ .
\label{cd}
\eea
In the remainder of this section, we shall drop the ``bar'' on the covariant derivatives as well as the $\Gamma$-matrices for simplicity in notation. Since the (\ref{cd}) contains gauge field, we adopt the spin-weighted harmonics $ _{s-\ft12}\eta^{(\ell)} $,  which are described in detail in appendix B, as the expansion basis. In this section we will need the harmonics for $s=0$, namely $ _{-\ft12} \eta^{(\ell)} $ which we will denote as $\eta^{(\ell)}$ for brevity in notation. These harmonics satisfy the relations,
\be
\eta_-^{(0)}=\eta\ ,\qquad
\eta^{(\ell)}_+ = \frac1{i\alpha\sqrt{\cell}}\nabla_n Y^{(\ell)}\sigma^n\eta\ ,\qquad
\eta^{(\ell)}_- = Y^{(\ell)}\eta,\qquad \ell\geq1\ ,
\ee
and have the following properties
\begin{alignat}{2}
 \sigma_3 {\eta_{\pm}^{(\ell)}}&=\pm{\eta_{\pm}^{(\ell)}}\ ,
 &\qquad\quad \sigma^n {D}_n{\eta_{\pm}^{(\ell)}}& =i\alpha\sqrt{\cell}\,{\eta_{\mp}^{(\ell)}}\ ,
\nn\\
[{D}_m, {D}_n]{\eta_-^{(\ell)}}&=0\ , &\qquad\quad [{D}_m,{D}_n]{\eta_+^{(\ell)}} &=i\alpha^2\epsilon_{mn}{\eta_+^{(\ell)}} ,
\nn\\
{D}^n{D}_n{\eta_-^{(\ell)}}&=-\alpha^2\cell{\eta_-^{(\ell)}}, &\qquad\quad {D}^n{D}_n{\eta_+^{(\ell)}}&=-\alpha^2(\cell-1){\eta_+^{(\ell)}}\ .
\end{alignat}
The Killing spinor $\eta$ also has the property $\sigma_3\eta=-\eta$. Furthermore, given that $\Gamma_7=\gamma_5\times \sigma_3$ (see Appendix A), the chirality property of a spinor in $6D$ correlates the $4D$ and $\sigma_3$ chiralities.

Since there is no gamma traceless and transverse spin-3/2 harmonics on the $S^2$, generically, the harmonic expansion are carried out as
\begin{eqnarray}
\hat{\psi}_{\mu} &=&\psi_{\mu-}^{(0)}\otimes{\eta^{(0)}}
+\sum_{\ell\geq1}\left(\psi_{\mu+}^{(\ell)}\otimes{\eta_+^{(\ell)}}
+\psi_{\mu-}^{(\ell)}\otimes{\eta_-^{(\ell)}}\right)\ ,
\nn\\
\hat{\psi}_m &=& \Gamma_m\psi^{(0)}_+\otimes{\eta^{(0)}}
+\Gamma_m\sum_{\ell\geq1}\left(\psi_-^{(\ell)}\otimes{\eta_+^{(\ell)}}
+\psi_+^{(\ell)}\otimes{\eta_-^{(\ell)}}\right)
\nn\\
 &&+\sum_{\ell\geq1}\left(\tilde\psi_+^{(\ell)}\otimes D_{\{m\}}{\eta_+^{(\ell)}}+\tilde\psi_-^{(\ell)}\otimes D_{\{m\}}{\eta_-^{(\ell)}}\right)\ ,
 \nn\\
 \hat{\chi}&=&\chi^{(0)}_+ \otimes{\eta^{(0)}}+ \sum_{\ell \geq1} \left(\chi_-^{(\ell)}\otimes{\eta_+^{(\ell)}}+\chi_+^{(\ell)}\otimes{\eta_-^{(\ell)}}\right)\ ,
\nn\\
\hat{\lambda} &=& \lambda^{(0)}_-\otimes \eta^{(0)}
+\sum_{\ell \geq1}\left( \lambda_+^{(\ell)} \eta_+^{(\ell)}
+\lambda_-^{(\ell)} \eta_-^{(\ell)}\right)\ ,
\end{eqnarray}
where $D_{\{m\}}$ is the gamma traceless covariant derivative and the $\pm$ subscripts denote chirality property under $\gamma_5$. Using the 6D linearized fermionic gauge symmetry (\ref{6dgs}), one can impose the following gauge condition
\be
\hat{\psi}_{\{m\}}=0\ ,
\label{fg}
\ee
where $\{m\}$ means $\Gamma$-traceless. As a consequence, the expansion takes the following simpler forms
\bea
\hat{\psi}_{\mu} &=&\psi_{\mu -}^{(0)}\otimes{\eta^{(0)}}
+\sum_{\ell\geq1}\left(\psi_{\mu+}^{(\ell)}\otimes{\eta_+^{(\ell)}}
+\psi_{\mu-}^{(\ell)}\otimes{\eta_-^{(\ell)}}\right)\ ,
\\
 \hat{\psi}_m &=& \Gamma_m\psi^{(0)}_+\otimes{\eta^{(0)}}
 +\Gamma_m\sum_{\ell\geq1}\left(\psi_-^{(\ell)}\otimes{\eta_+^{(\ell)}}
 +\psi_+^{(\ell)}\otimes{\eta_-^{(\ell)}}\right)\ ,
\\
\hat{\chi} &=& \chi^{(0)}_+\otimes{\eta^{(0)}}
+ \sum_{\ell \geq1}\left(\chi_-^{(\ell)}\otimes{\eta_+^{(\ell)}} +\chi_+^{(\ell)}\otimes{\eta_-^{(\ell)}}\right)\ ,
\\
\hat{\lambda} &=& \lambda^{(0)}_-\otimes{\eta^{(0)}}+\sum_{\ell \geq1} \left(\lambda_+^{(\ell)}\otimes{\eta_+^{(\ell)}}
+\lambda_-^{(\ell)}\otimes{\eta_-^{(\ell)}}\right)\ .
\eea
The gauge choice \eq{fg} does not fix all the gauge symmetries, we find the following residual symmetry transformations
\begin{itemize}
\item  Generated by $\hat{\epsilon}^{(0)}=\epsilon^{(0)}{\eta^{(0)}}$:
\begin{equation}\label{fs-1}
\delta\psi^{(0)}_\mu=\partial_{\mu}\epsilon^{(0)},
\end{equation}
\item Generated by $\hat{\epsilon}=\epsilon_+^{(1)}{\eta_+^{(1)}}$:

\begin{eqnarray}
\delta \psi_{\mu+}^{(1)}&=&\partial_{\mu}\epsilon_+^{(1)}
+\frac{i}{\sqrt 2}\,\alpha\,\epsilon_+^{(1)}\ ,
\\
\delta \lambda_+^{(1)}&=&ig\epsilon_+^{(1)}\ .
\label{fgs1}
\end{eqnarray}
\end{itemize}
We shall take into account these symmetries in the analysis of the spectrum below, where we treat the spin-3/2, spin-1/2 sectors separately.
\bigskip


\subsection*{\fbox{Spin-3/2 sector}}


\bigskip

Let us begin with the restriction $\ell\ge 1$. This sector contains only the gravitino fields which satisfy the following equations
\bea
\ell\ge 1:\quad     &&\pa\Big((\hat{\Box}_0+\alpha^2)-M^2\Big) P^{3/2}\psi_{\mu+}^{(\ell)}
    +i\alpha\sqrt{\cell}\Big((\hat{\Box}_0+\alpha^2)-M^2\Big)P^{3/2}\psi_{\mu-}^{(\ell)}=0\ ,
\nn \\
 &&i\alpha\sqrt{\cell}\Big((\hat{\Box}_0+\alpha^2)-M^2\Big)P^{3/2}\psi_{\mu+}^{(\ell)}
 -\pa\Big(\hat{\Box}_0-M^2\Big)P^{3/2}\psi_{\mu-}^{(\ell)}=0\ ,
\eea
where $P^{3/2}$ is the spin-3/2 projector operator defined in Appendix C.
Diagonalising the associated $2\times 2$ operator-valued matrix, we
find that the modes are annihilated by the partially-factorising
operator polynomial, of sixth order in $\pa$, given by
\be
\ell \ge 1:\qquad {\cal O}^{(3/2)} = {\cal O}_1 {\cal O}_2\ .
\label{spin1.5bb}
\ee
Next, consider the case of $\ell=0$. By choosing the gauge $\gamma^{\mu}\psi^{(0)}_{\mu}=0$, the gravitino equation can be written as
\be
\pa\left(\Box-M^2\right)\psi^{(0)\mu}=-\left(
\pa\partial^{\mu} - M^2\gamma^{\mu}\right)\Psi^{(0)}-2M^2\left( \gamma^{\mu\nu}\partial_{\nu}\psi^{(0)}
-i\gamma^{\mu\nu}\partial_{\nu}\chi^{(0)}\right)\ .
\ee
The solutions of above equation can be expressed as
$\psi^{(0)}_{\mu}=\psi'^{(0)}_{\mu}+\psi''^{(0)}_{\mu}$ where $\psi''^{(0)}_{\mu}$ is completely determined by $\psi^{(0)}$ and $\chi^{(0)}$ while $\psi'^{(0)}_{\mu}$ is the solution to the following equations modular gauge symmetry (\ref{fs-1})
\be
\ell=0:\qquad \pa(\Box-M^2)\psi'^{(0)}_{\mu}=0\ ,\qquad \gamma^{\mu}\psi'^{(0)}_{\mu}=0\ ,\qquad \partial^{\mu}\psi'^{(0)}_{\mu}=0\ .
\ee
It describes a massless and two massive gravitini.


\subsection*{\fbox{Spin-1/2 sector}}


The $\ell\ge 2$ sector consists of ten spin 1/2 fields $(\Lambda_+,\Psi_+,\Lambda_-,\Psi_-,$ $\psi_+,\psi_-,\chi_+,\chi_-,\lambda_+,\lambda_-)$, where $\Psi^{(\ell)}\equiv\partial^{\mu}\psi_{\mu}^{(\ell)}$ and
$\gamma^{\mu}\psi_{\mu}^{(\ell)}\equiv\Lambda^{(\ell)}$. The linearized equations describing their mixing are
\bea
0&=&
\pa\Lambda^{(\ell)}_{-}
-\Psi_+^{(\ell)} +i\alpha\sqrt{\cell}\psi_+^{(\ell)}
-i\alpha\sqrt{\cell}\Lambda_+^{(\ell)} +2\pa\psi_-^{(\ell)}
-2i\pa\chi_-^{(\ell)} +2\alpha\sqrt{\cell}\chi_+^{(\ell)}
-8gi\lambda_+^{(\ell)}\ ,
\label{spin.5-1}\w2
0&=&
\pa\Lambda_+^{(\ell)} -\Psi_-^{(\ell)}
+i\alpha\sqrt{\cell}\psi_-^{(\ell)}
+i\alpha\sqrt{\cell}\Lambda_-^{(\ell)} -2\pa\psi_+^{(\ell)}
+2i\pa\chi_+^{(\ell)} +2\alpha\sqrt{\cell}\chi_-^{(\ell)}
-8gi\lambda_-^{(\ell)}\ ,
\label{spin.5-2}\w2
0&=&
i g\Lambda_-^{(\ell)}
+2g\chi_-^{(\ell)} -\pa\lambda_+^{(\ell)}
-i\alpha\sqrt{\cell}\lambda_-^{(\ell)}\ ,
\label{spin.5-3}\w2
0&=&
2i g\psi_+^{(\ell)}
+2g\chi_+^{(\ell)}
+\pa\lambda_-^{(\ell)}
-i\alpha\sqrt{\cell}\lambda_+^{(\ell)}\ ,
\label{spin.5-4}\w2
0&=&
i\alpha\sqrt{\cell}\pa(\Box+M^2)\psi_+^{(\ell)}
-iM^2\alpha\sqrt{\cell}\pa\Lambda_+^{(\ell)}
+i\alpha\sqrt{\cell}\biggl(M^2-\alpha^2+\alpha^2\cell\biggr)\Psi_-^{(\ell)}
+2M^2\alpha\sqrt{\cell}\pa\chi_+^{(\ell)}
\nn\\
&&
-8igM^2\pa\lambda_+^{(\ell)}
+\alpha^2(\cell-1)\pa\Psi_+^{(\ell)}
+\alpha^2(2-\cell)\Box\psi_-^{(\ell)}\ ,
\label{spin.5-5}\w2
0&=&
-i\pa\biggl(\Box+M^2-\alpha^2)\biggr)\psi_-^{(\ell)}
-iM^2\pa\Lambda_-^{(\ell)}
+i\biggl(\alpha^2+\alpha^2\cell- M^2\biggr)\Psi_+^{(\ell)}
-2M^2\pa\chi_-^{(\ell)}
\nn\\
&&
-\alpha\sqrt{\cell}\left( \pa\Psi_-^{(\ell)} +\Box\psi_+^{(\ell)}\right)\ ,
\label{spin.5-6}\w2
0&=&
\pa\biggl(\Box+\alpha^2-\alpha^2\cell+2M^2\biggr)\Lambda_-^{(\ell)}
-\biggl(\Box+2\alpha^2-2\alpha^2\cell+2M^2\biggr)\Psi_+^{(\ell)}
+i\alpha\sqrt{\cell}(\Box+4M^2)\psi_+^{(\ell)}
\nn\\
&&+\pa\biggl(6M^2+2\alpha^2-\alpha^2\cell)\biggr)\psi_-^{(\ell)}
-i\alpha\sqrt{\cell}\biggl(\Box+\alpha^2-\alpha^2\cell +3M^2\biggr)\Lambda_+^{(\ell)}
-6iM^2\pa\chi_-^{(\ell)}
\nn\\
&&
+8\alpha M^2\sqrt{\cell}\chi_+^{(\ell)}
-32igM^2\lambda_+^{(\ell)}
+i\alpha\sqrt{\cell}\pa\Psi_-^{(\ell)}\ ,
\label{spin.5-7}\w2
0&=&
-\pa\biggl(\Box-\alpha^2\cell +2M^2)\biggr)\Lambda_+^{(\ell)}
+\biggl(\Box-2\alpha^2\cell +2M^2)\biggr)\Psi_-^{(\ell)} -i\alpha\sqrt{\cell}\biggl(\Box-4\alpha^2+4M^2)\biggr)\psi_-^{(\ell)}
\nn\\
&&+\pa(6M^2-\alpha^2\cell)\psi_+^{(\ell)}
-i\alpha\sqrt{\cell}\biggl(\Box+\alpha^2-\alpha^2\cell +3M^2\biggr)\Lambda_-^{(\ell)}
-6iM^2\pa\chi_+^{(\ell)}
-8\alpha M^2\sqrt{\cell}\chi_-^{(\ell)}
\nn\\
&&+i\alpha\sqrt{\cell}\pa\Psi_+^{(\ell)}\ ,
\label{spin.5-8}\w2
0&=& M^2\left(-\Lambda_-^{(\ell)}+2i\chi_-^{(\ell)}\right)
-\left(2\Box-\alpha^2\cell \right)\psi_-^{(\ell)}
+\pa\Psi_+^{(\ell)} +i\alpha\sqrt{\cell}\Psi_-^{(\ell)}
+i\alpha\sqrt{\cell}\pa\psi_+^{(\ell)}\ ,
\label{spin.5-9}\w2
0&=&
(\alpha^2-M^2)\Lambda_+^{(\ell)}
-2iM^2\chi_+^{(\ell)} +\left(2\Box+2\alpha^2- \alpha^2\cell \right)\psi_+^{(\ell)}
-i\alpha\sqrt{\cell}\Psi_+^{(\ell)}+\pa\Psi_-^{(\ell)}
+i\alpha\sqrt{\cell}\pa\psi_-^{(\ell)}\ .
\nn\\
&&
\eea
Diagonalising the associated $10\times 10$ operator-valued matrix, we
find that the modes are annihilated by the partially-factorising
operator polynomial, of eighteenth order in $\pa$, given by
\be
\ell \ge 2:\qquad {\cal O}^{1/2} = {\cal O}_1^5\, {\cal O}_4\ .
\label{spin.5b}
\ee
Next we consider the case of $\ell=1$.  In this case, one can use the fermionic shift symmetry \eq{fgs1} to eliminate $\psi_+^{(1)}$. Consequently we get 9 by 9 mixing and we find that the modes are annihilated by the partially-factorising operator polynomial, of fifteenth order in $\pa$, given by
\be
\ell=1:\qquad  {\cal O}^{1/2}  = {\cal O}_1^4|_{\ell=1}\, \pa\, {\cal O}_3\ ,
\ee
where ${\cal O}_3$ is defined in \eq{op3}. The factor $\pa$ demonstrates that there is massless spin-1/2 mode. This massless mode corresponds to linear combinations of $(\Lambda^{(1)}_+,\Psi^{(1)}_+,\Lambda^{(1)}_-,\Psi^{(1)}_-,\psi^{(1)}_-,\chi^{(1)}_+,\chi^{(1)}_-,\lambda^{(1)}_+,
\lambda^{(1)}_-)$ with mixing coefficients $(8,0,0,0,-2,-2i,0,0,1)$.

There remains the case of $\ell=0$. In this case, we have
\bea
\ell=0:\qquad 0&=&-\Psi^{(0)}-2\pa\psi^{(0)}+2i\pa\chi^{(0)}-8gi\lambda^{(0)}\ ,
\\
0&=& 2gi\psi^{(0)}+2g\chi^{(0)}+\pa\lambda^{(0)}\ ,
\\
0&=& (1+\frac{1}{2M^2}\Box)\Psi^{(0)}+3\pa\psi^{(0)}-3i\pa\chi^{(0)}\ ,
\\
0&=& -\Psi^{(0)}-\pa\left(1+\frac{1}{M^2} (\Box+\alpha^2)\right)\psi^{(0)}+2i\pa\chi^{(0)}-8gi\lambda^{(0)}\ .
\eea
Diagonalising the associated $4\times 4$ operator-valued matrix, we
find that the modes are annihilated by the partially-factorising
operator polynomial, of  seventh order in $\pa$, given by
\be
\ell=0:\qquad {\cal O}^{(1/2)} = {\cal O}_1|_{\ell=0}\,\slashed{\partial}\,\widetilde{\cal O}_2\ .
\ee
Thus, at the $\ell=0$ level, there is only one massless spin-1/2 modes given by $\Psi^{(0)}=0,\lambda^{(0)}=0,i\psi^{(0)}+\chi^{(0)}=0$.


\subsection{The supermultiplet structure and stability}


In arranging the full spectrum described above into a collection of supermultiplet structure, it is useful to recall that following massive supermultiplets:
\begin{alignat}{2}
&{\rm massive\ supergravity\ multiplet:} & \qquad (h_{\mu\nu}, A_\mu, \Psi_\mu)\ ,
\nn\\
&{\rm massive\ gravitino \ multiplet:} & \qquad (\psi_\mu, Z_\mu, \chi)\ ,
\nn\\
&{\rm massive\ vector multiplet\ multiplet:} & \qquad (A_\mu, \phi, \lambda)\ ,
\nn\\
&{\rm massive\ scalar\ multiplet:} & \qquad (Z, \psi)\ ,
\end{alignat}
where $A_\mu$ is a real and $Z_\mu$ is a complex vector, $\phi$ is a real and $Z$ is a complex scalar, and $\Psi_\mu$ is Dirac and  $\psi_\mu,\chi, \psi$ are Majorana. The Dirac gravitino can be written as $\Psi=\psi_{\mu+}^1 + \psi_{\mu-}^2$, where the two terms represent Weyl spinors that are independent of each other, and consequently $\Psi_\mu$ on-shell describes $8$ real degrees of freedom. The Majorana gravitino, on the other hand can be written as $\psi_\mu=\psi_{\mu+} + \psi_{\mu-}$ where $\psi_{\mu-}= (\psi_{\mu +})^*$. Thus, on shell $\psi_\mu$ describes $4$ real degrees of freedom. With this information at hand, we can now tabulate the supermultiplet structure of the full spectrum. It is convenient to do so by specifying the wave operators for different spin fields and consider the cases of $\ell=0$, $\ell=1$ and $\ell\ge 2$ separately. The results are given in Table 1, Table 2 and Table 3. \\
\begin{table}[ht]
\centering
\begin{tabular}{|c|c|c|c|c|}
  \hline
  $s=2$   & $s=3/2$  &  $s=1$   & $s=1/2$ & $s=0$  \\
  \hline\hline
  $\Box$    &  $\slashed{\partial}$    &      &         &         \\
  \hline
  $\Box-M^2$        &  $\Box-M^2$    & $\Box-M^2$   &     &         \\
  \hline
          &          & $\T2$    & $\T2$   & $\T2$    \\
  \hline
          &          &    & $\slashed{\partial}$  & $\Box^2$  \\
  \hline
          &          &    & $\1$ & $\1^4$      \\
  \hline
\end{tabular}
\caption{The spectrum of wave operators for $\ell=0$. The operator $\1$ is to be evaluated for $\ell=0$. There is one massless spin-2 and  one massless spin-0 multiplet, a massive spin-2 multiplet with mass $M$, two spin-0 multiplets with squared mass $m^2=M^2-\alpha^2$ and two massive spin-1 multiplets with mass$^2$ given in \eq{massT2}. }
\label{Table1}
\end{table}

\bigskip
\bigskip

\begin{table}[ht]
\centering
\begin{tabular}{|c|c|c|c|c|}
  \hline
  $s=2$   & $s=3/2$  &  $s=1$   & $s=1/2$ & $s=0$  \\
  \hline\hline
  $\2$    &  $\2$    &  $\2$    &         &         \\
  \hline
          &  $\1$    & $\1^4$   & $\1$    &         \\
  \hline
          &          & $\3$     & $\3$    & $\3$    \\
  \hline
          &          & $\1^2$   & $\1^2$  & $\1^2$  \\
  \hline
          &          & $\Box$   & $\slashed{\partial}$ &  \\
  \hline
          &          &          & $\1$  & $\1^4$  \\
  \hline
\end{tabular}
\caption{The spectrum of wave operators for $\ell=1$. The operators $\1$ and $\2$ are to be evaluated for $\ell=1$. There are two spin-2 multiplets with squared masses given in \eq{mass2} for $\ell=1$, two spin-3/2, two spin-1 multiplets and two spin-0 multiplets with squared mass $m^2=M^2+\alpha^2$,  three spin-1 multiplets with squared masses given by the roots of the polynomial given in \eq{mass3} and a massless vector multiplet.}
\label{Table2}
\end{table}

\begin{table}[ht]
\centering
\begin{tabular}{|c|c|c|c|c|}
  \hline
  $s=2$   & $s=3/2$  &  $s=1$   & $s=1/2$ & $s=0$   \\
  \hline\hline
  $\2$    &  $\2$    &  $\2$    &         &         \\
  \hline
          & $\1$     & $\1^4$   & $\1$    &         \\
  \hline
          &          & $\4$     & $\4$    & $\4$    \\
  \hline
          &          & $\1^2$   & $\1^2$  & $\1^2$  \\
  \hline
          &          &          & $\1^2$  & $\1^8$  \\
  \hline
\end{tabular}
\caption{The spectrum of wave operators for $\ell\ge 2$. For each integer $\ell$, there are two spin-2 multiplets with squared mass $m^2_\pm(\ell)$ given in \eq{mass2}, two spin-3/2, two spin-1 multiplets and four spin-0 multiplets with squared mass $m^2(\ell)$ given in \eq{mass1}, and four spin-1 multiplets with squared masses given by the roots of the polynomial given in \eq{mass4}. }
\label{Table3}
\end{table}

\bigskip
\bigskip
\bigskip
\bigskip

   Having established the full spectrum of states in the four-dimensional
theory, we may now examine the question of stability, which is governed by
the mass values for the massive fields.  We begin with the
$\ell=0$ level, given in Table 1. In addition to the
 massless graviton and scalar multiplets, there are massive graviton, vector
and scalar multiplets at this level.  The massive graviton multiplet has
$m^2=M^2$, the scalar multiplet has $m^2=M^2-\alpha^2$, and the vector
multiplet has masses given by (\ref{massT2}).  These imply, respectively,
that stability requires $M^2>0$, $M^2>\alpha^2$ and $M^2\ge  (
5+2\sqrt6)\alpha^2 \approx 9.89898 \alpha^2$.

    At the level $\ell=1$, in addition to the massless vector multiplet,
there are massive graviton, gravitino, vector and scalar multiplets. The
gravitino multiplet, two of the massive vector multiplets and the scalar
multiplet have masses given by the operator ${\cal O}_1$, implying
$m^2=M^2+\alpha^2$, which therefore impose no new conditions.  The
massive graviton multiplet with mass operator ${\cal O}_2$ has
$m^2$ given by equation (\ref{mass2}) with $\ell$ set equal to 1.  This
implies $m^2= (M^2+4\alpha^2\pm \sqrt{M^4+8\alpha^4})/2$, and hence gives
no further restriction.  There remains the massive vector multiplet
with mass operator ${\cal O}_3$ given in (\ref{op3}).  This gives
a cubic polynomial in $m^2$, and we find that this has three real
(and positive) roots for $m^2$ provided that $\mu\equiv M^2/\alpha^2$
satisfies the condition
\be
4\mu^4 - 64\mu^3 + 153 \mu^2 - 26\mu - 139\ge0\,.
\ee
This implies we must have $\mu\ge \mu_{\rm min}$, where $\mu_{\rm min}
\approx 13.1425$. In other words, at level $\ell=1$ stability requires
that $M^2$ should exceed approximately $13.1425 \alpha^2$.

   For levels $\ell\ge 2$, the multiplets are given in Table 3.  The
gravitino, vector and scalar multiplets with mass operator ${\cal O}_1$
have $m^2$ given in equation (\ref{mass1}), and these are always positive
for all values of $\ell\ge 2$.  Likewise, for the graviton multiplet
with mass operator ${\cal O}_2$, $m^2$, given in (\ref{mass2}), is positive
for all $\ell\ge2$.  There remains the vector multiplet with mass operator
${\cal O}_4$.  This leads to a quartic polynomial in $m^2$, which can be
read off from (\ref{1234}) and (\ref{boxhat}).  One can show that this
polynomial necessarily has at least two real roots, which are positive,
and that if the
four roots for $m^2$ are real then they are also positive.  The
condition for having four real roots is that a rather complicated discriminant
of sixth order in $\mu=M^2/\alpha^2$ should be positive.  This discriminant
also depends on the level $\ell$.  For a few representative values of $\ell$,
we find the requirement $\mu\ge \mu_{\rm min}(\ell)$:

\begin{table}[ht]
\centering
\begin{tabular}{|c|c|c|c|c|c|c|c|}
  \hline
$\ell=$ & 2 & 3 & 5 & 10 & 100 & 1000 & 10000\\
\hline
$\mu_{\rm min}\approx$ & 16.9381& 20.869 & 28.8614 & 49.0439 & 414.215&
                  4067.07 & 40595.8\\
\hline
\end{tabular}
\caption{Minimum values of $\mu=M^2/\alpha^2$ necessary to achieve real
positive mass-squared values for the ${\cal O}_4$ vector multiplet at level
$\ell$.}
\label{Table4}
\end{table}

\bigskip
\bigskip
\bigskip
\bigskip

In the limit of large $\ell$, we find that to leading order,
$\mu_{\rm min}(\ell)$ grows linearly with $\ell$, with
\be
\mu_{\rm min}(\ell) \sim  4.05874 \, \ell +\cdots\label{mul}
\ee
This implies that for any given ratio $\mu=M^2/\alpha^2$, there is a
a critical level $\ell_{\rm max}$ beyond which the Kaluza-Klein tower
must be truncated in order not to have modes with complex masses, which
would be associated with instabilities.

\section{Spectrum in Non-Supersymmetric $\rm{Minkowski}_4\times S^2$ Background}


\subsection{Non-supersymmetric $\rm{Minkowski}_4\times S^2$ background}


In addition to supersymmetric vacuum solution discussed in previous section, the theory \cite{Coomans:2012ax} also possesses non-supersymmetric $\rm{Minkowski}_4\times S^2$ vacua when $M^2=\alpha^2$, with the curvature and flux given by
\begin{alignat}{3}
 \bar{R}_{\mu\nu\lambda\rho} &=0 \ ,&\qquad  \bar{R}_{mn} &= \alpha^2\bar{g}_{mn}\ , \qquad \bar{L} &=1\ , \nonumber\\
  \bar{F}_{\mu\nu} &=0\ , & \qquad  \bar{F}_{mn} &= 4q g\epsilon_{mn}\ , \qquad &  \nonumber\\
  \bar{G}_{\mu\nu} &=0\ , & \qquad \bar{G}_{mn} &= -q\alpha^2\epsilon_{mn}\ , \qquad &
\label{nonsusy}
\end{alignat}
where $q$ plays the role of monopole charge and is quantized to be $q=0,\pm1,\pm2\ldots$. The supersymmetric vacua correspond to $q=\pm1$.


\subsection{Bosonic sector}


We will perform a similar spectrum analysis around the non-supersymmetric background.
The harmonic expansion (\ref{HE}) for the uncharged fields after the gauge fixing is still valid, and the residual
gauge symmetries are almost the same except that some terms related to background flux should be
multiplied by the monopole charge. We present the results for the spectrum below. While we shall use the same notation for operators such as $\hat\Box_0$ and others, it is understood that they are to be evaluated for $M^2=\alpha^2$.


\subsection*{\fbox{Spin-2 sector}}


\bigskip

The equations of motion satisfied by graviton for $\ell\ge 1$ is
\be
\ell \ge 1:\qquad     ( \hat\Box_0^2-\alpha^2\hat\Box_0-\alpha^4\, \cell)(\cP^{2} h)^{(\ell)}_{\mu\nu}=0\ ,
\ee
describing massive gravitons with square masses
\be
m^2_\pm (\ell) = \frac12 \alpha^2\left(1+c_\ell \pm \sqrt{1+4c_\ell}\right)\ .
\ee
For $\ell=0$, the linearized field equation is
\be
\ell=0:\qquad (\Box-\alpha^2)R^{L(0)}_{\mu\nu}=-\alpha^2\partial_{\mu}\partial_{\nu}S^{(0)}-\alpha^4\eta_{\mu\nu}S^{(0)}
+\partial_{\mu}\partial_{\nu}(\Box+\alpha^2)S^{(0)}\ ,
\ee
where $S^{(0)}=\phi^{(0)}+N^{(0)}$. It describes a massless graviton and massive graviton with squared mass $m^2=\alpha^2$.

\bigskip
\bigskip

\subsection*{\fbox{Spin-1 sector}}


\bigskip

For $\ell\ge 2$ the mixing among the five vector fields $(k^{T}_{\mu},a^{T}_{\mu},v^{T}_{\mu}$ $,b_{\mu\nu}^{T},b^{T}_{\mu})$ now have the following form
\bea
\ell\ge 2:\quad
0&=&(2\cell\alpha^4-\hat{\Box}_0^2)k^{T(\ell )}_{\mu}-4g\alpha^2q a^{T(\ell )}_{\mu}-\frac{\alpha^2}{2}(4 q v^{T(\ell )}_{\mu}
-\epsilon_{\mu}^{~\nu\lambda\rho}\partial_{\nu}b_{\lambda\rho}^{T(\ell )})\ ,
\w2
0&=&
(\alpha^2\,\Box-\hat{\Box}^2_0)b^{T(\ell )}_{\mu\nu}
-4q g\alpha^2\star F^{(\ell)}_{\mu\nu}(a)+\alpha^4\cell(\star F^{(\ell)}_{\mu\nu}(k)
-\star F^{(\ell)}_{\mu\nu}(b))\ ,
\w2
0&=& \hat{\Box}_0^2b^{T(\ell )}_{\mu}
+\ft{1}{2}\alpha^2\epsilon_{\mu}^{~\nu\lambda\rho}\partial_{\nu}b_{\lambda\rho}^{T(\ell )}\ ,
\w2
0&=& (\hat{\Box}_0+\alpha^2)a^{T(\ell )}_{\mu} -4q g\alpha^2\cell k^{T(\ell )}_{\mu}+4gv_{\mu}^{T(\ell )}-2q g\epsilon_{\mu}^{~\nu\lambda\rho}\partial_{\nu}b_{\lambda\rho}^{T(\ell )}\ ,
\w2
0&=& (\hat{\Box}_0-\alpha^2)v^{T(\ell )}_{\mu}+\alpha^4q\cell k^{T(\ell )}_{\mu}-2g\alpha^2 a^{(\ell)}_{\mu}\ ,
\label{mix1}
\eea
Diagonalising the associated $5\times 5$ operator-valued matrix, we find that the modes are annihilated
by the partially-factorising operator polynomial, of eighth order in $\hat\Box_0$, given by $\hat{\Box}_0^2\,{\cal O}_6$.
The explicit form of ${\cal O}_6$ can be obtained straightforwardly from \eq{mix1}, and one finds that it is symmetric
under $q\rightarrow-q$ meaning that vector spectrum is symmetric under the sign change of monopole charge.
Of the remaining vectors $((\cP^{1}h)_{\mu\nu}, \partial^\mu b_{\mu\nu})$ are annihilated by $\hat{\Box}_0$.
Thus, apart from the charged vectors which will be treated separately below, the total wave operator for $\ell\ge 2$ is given by
\be
\ell \ge 2:\qquad {\cal O}^{(1)} = \hat{\Box}_0^4|_{M^2=\alpha^2}\,{\cal O}_6\ ,
\ee
implying four massive vectors with squared masses $m^2=\alpha^2 c_\ell$, and six massive vectors whose
squared masses $m^2$ correspond to the roots ${\cal O}_6$ in which $\Box$ is to be replaced by $m^2$.

In the case of $\ell=1$, again excluding the charged vector, we find that a massless vector appears since for
$\cell=2$, the operator  ${\cal O}_6 $ factorizes as ${\cal O}_6= \Box\, {\cal O}_5$, and the total wave operator
becomes
\be
\ell=1:\qquad {\cal O}^{(1)} = \hat{\Box}_0^4|_{\ell=1}\,\Box\, {\cal O}_5\ .
\label{pp3}
\ee
  The massless vector is composed from a linear combination of $(k^{T(1)}_{\mu},a^{T(1)}_{\mu},v^{T(1)}_{\mu},b_{\mu\nu}^{T(1)},b^{T(1)}_{\mu})$ with mixing coefficients $\left(\ft{2}{1+q^2},-\ft{8q g}{1+q^2},\ft{2q\alpha^2}{1+q^2},1,0\right)$.

In the uncharged vector sector, there remains the case of $\ell=0$, for which the relevant vector fields are $(b_{\mu\nu}^{T(0)},a^{T(0)}_{\mu},v^{T(0)}_{\mu})$. Upon diagonalising the associated $3\times 3$ operator-valued matrix, we find that the modes are annihilated
by the following partially-factorising operator polynomial
\be
\ell=0: \qquad  {\cal O}^{(1)}= \Box(\Box-\alpha^2)(\Box^2+2\alpha^4q^2)\ .
\label{vecmass0}
\ee
As before, we find that the would-be massless modes annihilated by $\Box$ is eaten by the two form. Thus there are no massless vector modes at $\ell=0$.

Finally, we turn to the treatment of the complex vector $\hat{z}_{\mu}$. This field $\hat{z}_{\mu}$ is expanded in terms of charge ``-$q$'' scalar harmonics starting from $\ell=|q|$ as follows:
\be
{\hat z}_\mu = \sum_{\ell\ge q} z^{(\ell)}_\mu\, {}_{-q}{Y}^{(\ell)}\ ,
\ee
The resulting linearized field equation is
\be
\ell\geq |q|:\qquad \left(\Box-\alpha^2\cell+\alpha^2q^2-M^2\right)\, z^{T(\ell)}_{\mu}=0\ .
\ee

\bigskip


\subsection*{\fbox{Spin-0 sector}}


\bigskip

For $\ell\ge 2$, the equations describing the mixing between $(\phi, N, \varphi, \tilde{\varphi},a, v)$ take the following form
\bea
\ell\ge 2:\quad  0&=&2(\hat{\Box}_0+\alpha^2)\phi^{(\ell)}+(2\hat{\Box}_0+2\alpha^2+\alpha^2\cell)N^{(\ell)}+3\hat{\Box}_0\varphi^{(\ell)}
    -\alpha^2\cell \tilde{\varphi}^{(\ell)}\ ,
\w2
0&=&\alpha^2(3\varphi^{(\ell)}+2\phi^{(\ell)}-4gq a^{(\ell)}-2q v^{(\ell)})
  +\alpha^2(1-\cell)\tilde{\varphi}^{(\ell)}+\Box\,N^{(\ell)}\ ,
\w2
0&=&2\alpha^2\phi^{(\ell)}+(2\hat{\Box}_0+2\alpha^2+\alpha^2\cell)N^{(\ell)}-\Box\tilde{\varphi}^{(\ell)}\ ,
\w2
0&=& (\hat{\Box_0}+\alpha^2)a^{(\ell)}+4gq N^{(\ell)}-2gq (3\varphi^{(\ell)}+\tilde{\varphi}^{(\ell)})+4gv^{(\ell)}\ ,
\w2
0&=&(\hat{ \Box}_0-\alpha^2)v^{(\ell)}-\alpha^2 q N^{(\ell)}+\frac{\alpha^2}{2}q(3\varphi^{(\ell)}+\tilde{\varphi}^{\ell})
-2g\alpha^2 a^{(\ell)}\ ,
\w2
0&=&(\alpha^2\Box-\hat{\Box}^2_0)\varphi^{(\ell)}+2\alpha^4\phi^{(\ell)}+\alpha^4(2-\cell)N^{(\ell)}-4q g\cell\alpha^4 a^{(\ell)}-2\cell\alpha^4q v^{(\ell)} \ .
\eea
Diagonalising the associated $6\times 6$ operator-valued matrix, we
find that the modes are annihilated by the partially-factorising
operator polynomial, of seventh order in $\hat\Box_0$, given by ${\cal O}_4\hat{\Box}_0^3|_{M^2=\alpha^2}$
Of the remaining scalars,  $(\partial^{\mu}k_{\mu},\partial^\mu b_{\mu},\partial^\mu a_{\mu},\partial^\mu v_{\mu},\tilde{v})$, three of them namely $(\partial^{\mu}k_{\mu},\partial^\mu b_{\mu},\tilde{v}) $ are annihilated by ${\hat\Box}_0$, and the remaining two are determined in terms of them.  Thus, in total, apart from the complex scalars which will be treated separately below, the wave operator for the scalar fields is given by
\be
{\cal O}^{(0)} = {\hat\Box}_0^6\,{\cal O}_4\ .
\ee

Next, consider the case $\ell=1$. Utilizing the residual symmetry (\ref{bs-2}) and (\ref{bs-3}), one can eliminate $N^{(1)}$ and $\partial^\mu k_{\mu}^{(1)}$. The mass operator coming from the mixing among $(\phi,\varphi, \tilde{\varphi},a, v)$ takes the form $\hat\Box_0^2|_{\ell=1}{\cal O}_3$. Taking into account $\partial^\mu b_\mu^{(1)}$ and $\widetilde v^{(1)}$, the total wave operator, again, excluding the complex scalar sector, is given by
\be
\ell=1:\qquad {\cal O}^{(0)} = \hat{\Box}_0^4\,{\cal O}_3\ .
\ee

There remaining the case of $\ell=0$. In this case, the relevant scalar fields are $(\phi^{(0)},N^{(0)},b^{(0)})$, and they satisfy the following equations respectively
\bea
\ell=0:\quad &&(\Box^2+2\alpha^4)S^{(0)}=0\ ,\qquad S^{(0)}=\phi^{(0)}+N^{(0)}\ ,
\nn\\
&& \Box^2b^{(0)}=0\ ,
\nn\\
&&\Box^2N^{(0)}= 0\ .
\eea
Thus besides two massless modes, we also have modes with linear time coordinate dependence.
Finally, we discuss the complex scalars originating from $\hat{z}_m$. For positive monopole charge, the harmonic expansion of $\hat{z}_m$ is given by
\be
\hat{z}_m=z^{(q-1)} {}_{-q}V_m^{(q-1)}+z^{(q)} {}_{-q}V_m^{(q)}
+\sum_{\ell>q}(z^{(\ell)}D_m{}_{-q}Y^{(\ell)}+\tilde{z}^{(\ell)}\epsilon_m^{~n}D_n{}_{-q}Y^{(\ell)})\ .
\ee
Thus we have
\begin{alignat}{3}
&&\ell>q &:\qquad z^{(\ell)}\ ,\ \tilde{z}^{(\ell)}   &\qquad \mbox{with} &\qquad m^2=\alpha^2\cell-\alpha^2q+M^2\ ,
\nn\\
&&\ell=q &:\qquad z^{(q)}\ , &\qquad\mbox{with} &\qquad m^2=\alpha^2q+M^2\ ,
\nn\\
&&\ell=q-1 &:\qquad z^{(q-1)}\ , &\qquad\mbox{with} &\qquad m^2=M^2-\alpha^2q\ .
\end{alignat}


\subsection{Fermionic sector}


The analysis of the fermionic spectrum in a non-supersymmetric background (\ref{nonsusy}) is more subtle than that in supersymmetric background. Since the non-supersymmetric background do not posses Killing spinor, we will use
spin-weighted harmonics $ _{s-\ft12}\eta^{(\ell)}$, described in detail in appendix B, as basis of expansion. For brevity, we shall use the notation
\be
_{s-\ft12}\eta^{(\ell)} \equiv \tilde\eta^{(\ell)}\ ,
\label{nota1}
\ee
where $s=\ft{1}{2}(1-q)$. These harmonics satisfy the relations
\bea
&&\tilde{\eta}^{(\ell)}_+=
                      \eta_+ \left(_{s-1}Y^{(\ell)} \right)\ , \qquad   \tilde{\eta}^{(\ell)}_- = \eta_- \left( _{s}Y^{(\ell)} \right)\ ,
\\
&& (\frac{d}{d\theta}+m\csc\theta+s\cot\theta){}\left(_{s}Y^{(\ell)}\right)
=\sqrt{(\ell+s)(\ell+1-s)}{}\left(_{s-1}Y^{(\ell)}\right)\ ,\\
&&(\frac{d}{d\theta}-m\csc\theta-(s-1)\cot\theta){}\left(_{s-1}Y^{(\ell)}\right)
=-\sqrt{(\ell+s)(\ell+1-s)}{} \left(_{s}Y^{(\ell)}\right)\ .
\eea
The lowest level would have definite chirality when $\ell=-s$ for $q>0$ and $\ell=s-1$ for $q<0$. The spin weighted harmonics satisfy the following properties
\begin{alignat}{3}
\sigma_3\tilde{\eta}_{\pm}^{(\ell)} &=\pm\tilde{\eta}_{\pm}^{(\ell)}\ , &\qquad
\sigma^n D_n\tilde{\eta}_{\pm}^{(\ell)} &=i\alpha\sqrt{\tilde{\cell}}\,\tilde{\eta}_{\mp}^{(\ell)}\ ,
\nn\\
[D_m, D_n]\tilde{\eta}_-^{(\ell)}  &=-is\alpha^2\epsilon_{mn}\tilde{\eta}_-^{(\ell)}\ ,&\qquad [D_m,D_n]{\tilde{\eta}}_+^{(\ell)} &=i(1-s)\alpha^2\epsilon_{mn}{\eta_+^{(\ell)}}\ ,
\nn\\
D^nD_n\tilde{\eta}_-^{(\ell)} &=-\alpha^2(\tilde{\cell}-s)\tilde{\eta}_-^{(\ell)}\ ,&\qquad
D^nD_n\tilde{\eta}_+^{(\ell)} &= \alpha^2 (1-s-\tilde{\cell})\tilde{\eta}_+^{(\ell)}\ ,
\end{alignat}
where
\be
\tilde{\cell}=\biggl(\cell-s(s-1)\biggr)\ .
\ee

The harmonic expansion for 6D spin-$1/2$ fields follows the same procedure as in supersymmetric case by using the spin weighted harmonics, while it is more subtle when expanding the 6D gravitini.



It can be checked that the linearized equations have the following discreet symmetry
\begin{alignat}{3}
q\rightarrow-q\ , &\qquad \psi_{\mu+}\rightarrow-\psi_{\mu-}
 ,& \qquad\psi_{\mu-}\rightarrow\psi_{\mu+}\ ,
\nn\\
\psi_-\rightarrow\psi_+\ , &\qquad\psi_+\rightarrow-\psi_-\ , &\qquad
\chi_-\rightarrow\chi_+\ ,
\nn\\
\chi_+\rightarrow-\chi_-\ , &\qquad\lambda_+\rightarrow-\lambda_-\ ,
& \qquad\lambda_-\rightarrow\lambda_+\ ,
\end{alignat}
which implies that the spectrum keeps the same under the sign change of monopole charge. In the following, we will focus on the case with positive monopole charge and use $|s|$ to denote $\ft12(q-1)$.

%


\subsection*{\fbox{Spin-3/2 Sector}}


The gravitini satisfy
\bea
\ell\geq |s|+1:\qquad &&\pa\Big(\Box+\alpha^2|s|-\alpha^2\tilde{\cell}\Big)\psi_{\mu+}^{(\ell)}
+i\alpha\sqrt{\tilde{\cell}}\Big(\Box-\alpha^2\tilde{\cell}\Big)\psi_{\mu-}^{(\ell)}=0\ ,
\nn \\
&&i\alpha\sqrt{\tilde{\cell}}\Big(\Box-\alpha^2\tilde{\cell}\Big)\psi_{\mu+}^{(\ell)}
-\pa\Big(\Box+(|s|+1-{\tilde c}_\ell)\alpha^2\Big)\psi_{\mu-}^{(\ell)}=0\ .
\label{ge2}
\eea
Diagonalising the associated $2\times 2$ operator-valued matrix, we
find that the modes are annihilated by the partially-factorising
operator polynomial, of third order in $\Box$, given by
\be
{\cal O}^{(3/2)} = \tilde{{\cal O}}_3\ ,
\label{spin1.5b}
\ee
where the explicit form of $\tilde{{\cal O}}_3$ can be deduced from \eq{ge2}.

Next, we consider the case of $\ell=|s|$. In this case, we find that the quadratic action for the lowest level fermionic fields is proportional to
\bea
  {\cal L}^{(2)}&\propto& -i\bar{\chi}\gamma^{\mu\nu}\partial_{\mu}\psi_{\nu}-i\bar{\psi}_{\mu}\gamma^{\mu\nu}\partial_{\nu}\chi
  +2i\bar{\chi}\pa\psi-2i\bar{\psi}\pa\chi+2\bar{\chi}\pa\chi\nn\\
  &&-8g\bar{\chi}\lambda-8g\bar{\lambda}\chi-4\bar{\lambda}\pa\lambda-4i g|s|\bar{\lambda}\gamma^{\mu}\psi_{\mu}
  -4i g|s|\bar{\psi}_{\mu}\gamma^{\mu}\lambda\nn\\
 && -8g(1+|s|)i\bar{\lambda}\psi+8g(1+|s|)i\bar{\psi}\lambda
  -\ft{1}{2}\bar{\psi}_{\mu}\gamma^{\mu\nu\lambda}\partial_{\nu}\psi_{\lambda}+\bar{\psi}\pa\psi\nn\\
  &&+\bar{\psi}_{\mu}\gamma^{\mu\nu}\partial_{\nu}\psi-\psi\gamma^{\mu\nu}\partial_{\mu}\psi_{\nu}
  -\frac{1}{\alpha^2}\biggl(\ft{1}{4}\bar{\psi}_{\mu\nu}\pa\psi^{\mu\nu}-\bar{\psi}\pa\Box\psi\nn\\
 && +\ft{1}{2}|s|\alpha^2\bar{\psi}_{\mu}\pa\psi^{\mu}-|s|\alpha^2\bar{\psi}\partial_{\mu}\psi^{\mu}
  +|s|\alpha^2\bar{\psi}_{\mu}\partial^{\mu}\psi-(1+2|s|)\alpha^2\bar{\psi}\pa\psi\biggr)\ .
\eea
Unlike the supersymmetric case, we see the appearance of the terms $\bar{\lambda}\gamma^{\mu}\psi_{\mu}$,
$\bar{\psi}_{\mu}\pa\psi^{\mu}$ and $\bar{\psi}_{\mu}\partial^{\mu}\psi$ which break the fermionic gauge symmetry. The homogeneous solutions for gravitini satisfy
\be
\pa(\Box-\alpha^2-|s|\alpha^2)\psi_{\mu}^{(|s|)}=0,\qquad \gamma^{\mu}\psi^{(|s|)}_{\mu}=0,\qquad \partial^{\mu}\psi_{\mu}^{(|s|)}=0.
\label{gravitinomassless}
\ee
Thus, due to the lack of fermionic gauge symmetry, the longitudinal mode
$\psi_{\mu}\propto\,p_{\mu}e^{ipx}$ with $p^2=0$ becomes a dynamical degree of freedom.


\subsection*{\fbox{Spin-1/2 Sector}}


The $\ell\geq|s|+2$ sector consists of ten spin-1/2 fields $(\Lambda_+,\Psi_+,\Lambda_-,\Psi_-,$ $\psi_+,\psi_-,\chi_+,\chi_-,\lambda_+,\lambda_-)$. The linearized equations describing their mixing are
\bea
0&=&\pa\Lambda^{(\ell)}_{-}-\Psi_+^{(\ell)}+i\alpha\sqrt{\tilde{\cell}}\psi_+^{(\ell)}
-i\alpha\sqrt{\tilde{\cell}}\Lambda_+^{(\ell)}+2\pa\psi_-^{(\ell)}
-2i\pa\chi_-^{(\ell)}+2\alpha\sqrt{\tilde{\cell}}\chi_+^{(\ell)}-8gi\lambda_+^{(\ell)}\ ,
\label{1}
\\
0&=&\pa\Lambda_+^{(\ell)}-\Psi_-^{(\ell)}+i\alpha\sqrt{\tilde{\cell}}\psi_-^{(\ell)}
+i\alpha\sqrt{\tilde{\cell}}\Lambda_-^{(\ell)}-2\pa\psi_+^{(\ell)}
+2i\pa\chi_+^{(\ell)} +2\alpha\sqrt{\tilde{\cell}}\chi_-^{(\ell)}-8gi\lambda_-^{(\ell)}\ ,
\label{2}
\\
0&=&  g(1+|s|)i\Lambda_-^{(\ell)}+2g\chi_-^{(\ell)}-\pa\lambda_+^{(\ell)}
-i\alpha\sqrt{\tilde{\cell}}\lambda_-^{(\ell)}\ ,
\label{3}
\\
0&=&2i g|s|\psi_+^{(\ell)}+2g\chi_+^{(\ell)}+\pa\lambda_-^{(\ell)}
-i\alpha\sqrt{\tilde{\cell}}\lambda_+^{(\ell)}\ ,
\label{4}
\\
0&=&i\sqrt{\tilde{\cell}}\pa(\Box+\alpha^2+|s|\alpha^2)\psi_+^{(\ell)}
-i\alpha^2\sqrt{\tilde{\cell}}\pa\Lambda_+^{(\ell)}
+i\alpha^2\tilde{\cell}^{3/2}\Psi_-^{(\ell)}+2\alpha^2\sqrt{\tilde{\cell}}\pa\chi_+^{(\ell)}
\nn\\
&& -8g(1+|s|)\alpha\,i\pa\lambda_+^{(\ell)}+\alpha(\tilde{\cell}-1-|s|)\pa\Psi_+^{(\ell)}
+\alpha(2+2|s|-\tilde{\cell})\Box\psi_-^{(\ell)}\ ,
\label{5}
\\
0&=& i\sqrt{\tilde{\cell}}\pa(\Box-|s|\alpha^2)\psi_-^{(\ell)}
+i\alpha^2\sqrt{\tilde{\cell}}\pa\Lambda_-^{(\ell)}
-i\alpha^2\tilde{\cell}^{3/2}\Psi_+^{(\ell)}+2\alpha^2\sqrt{\tilde{\cell}}\pa\chi_-^{(\ell)}
+8ig|s|\alpha \pa\lambda_{-}^{(\ell)}
\nn\\
&&+\alpha(\tilde{\cell}+|s|)\pa\Psi_-^{(\ell)}
+\alpha(\tilde{\cell}+2|s|)\Box\psi_+^{(\ell)}\ ,
\label{6}
\\
0&=&\pa(\Box+3\alpha^2+|s|\alpha^2-\alpha^2\tilde{\cell})\Lambda_-^{(\ell)}
-(\Box+4\alpha^2+2|s|\alpha^2-2\alpha^2\tilde{\cell})\Psi_+^{(\ell)}
+i\alpha\sqrt{\tilde{\cell}}(\Box+4\alpha^2+4|s|\alpha^2)\psi_+^{(\ell)}
\nn\\
&&+\alpha^2\pa(8+2|s|-\tilde{\cell})\psi_-^{(\ell)}
-i\alpha\sqrt{\tilde{\cell}}(\Box+4\alpha^2-\alpha^2\tilde{\cell})\Lambda_+^{(\ell)}
-6i\alpha^2\pa\chi_-^{(\ell)}+8\alpha^3\sqrt{\tilde{\cell}}\chi_+^{(\ell)}
\nn\\
&&-32ig\alpha^2\,(1+|s|)\lambda_+^{(\ell)} +i\alpha\sqrt{\tilde{\cell}}\pa\Psi_-^{(\ell)}\ ,
\label{7}
\\
0&=&-\pa(\Box+2\alpha^2-|s|\alpha^2-\alpha^2\tilde{\cell})\Lambda_+^{(\ell)}
+(\Box+2\alpha^2-2|s|-2\alpha^2\tilde{\cell})\Psi_-^{(\ell)}- i\alpha\sqrt{\tilde{\cell}}(\Box-4|s|\alpha^2)\psi_-^{(\ell)}
\nn\\
&&+\alpha^2\pa(6-2|s|-\tilde{\cell})\psi_+^{(\ell)}
-i\alpha\sqrt{\tilde{\cell}}(\Box+4\alpha^2-\alpha^2\tilde{\cell})\Lambda_-^{(\ell)}
-6i\alpha^2 \pa\chi_+^{(\ell)}-8\alpha^3\sqrt{\tilde{\cell}}\chi_-^{(\ell)}
\nn\\
&&-32ig|s|\alpha^2 \lambda_-^{(\ell)}+i\alpha\sqrt{\tilde{\cell}} \pa\Psi_+^{(\ell)}\ ,
\label{8}
\\
0&=&(1+|s|)\alpha^2\Lambda_-^{(\ell)}-2i\alpha^2 \chi_-^{(\ell)}
+(2\Box-2 |s|\alpha^2-\alpha^2\tilde{\cell})\psi_-^{(\ell)}
-\pa\Psi_+^{(\ell)}-i\alpha\sqrt{\tilde{\cell}}\Psi_-^{(\ell)}
-i\alpha\sqrt{\tilde{\cell}}\pa\psi_+^{(\ell)}\ ,
\label{ 9}
\nn\\
&& \\
0&=&|s|\alpha^2\Lambda_+^{(\ell)}-2i\alpha^2 \chi_+^{(\ell)}
+(2\Box+2\alpha^2+2|s|\alpha^2-\alpha^2\tilde{\cell})\psi_+^{(\ell)} -i\alpha\sqrt{\tilde{\cell}}\Psi_+^{(\ell)}+\pa\Psi_-^{(\ell)}
+i\alpha\sqrt{\tilde{\cell}}\pa\psi_-^{(\ell)}\ .
\nn\\
\eea
Diagonalising the associated $10\times 10$ operator-valued matrix, we
find that the modes are annihilated by the partially-factorising
operator polynomial, of ninth order in $\Box$, given by
\be
\ell \ge |s|+2:\qquad {\cal O}^{(0)} = \hat{\Box}_0^3\,\tilde{{\cal O}}_6\ ,
\ee
where the explicit form of the operator ${\cal O}_6$ can be determined from the linearized spin 1/2 field equations listed above.

\fbox{$\ell=|s|+1$}\\

At this level, since $D_m{\eta_+^{(|s|+1)}}=\frac{i}{2}\alpha\sqrt{\tilde{\cell}}\sigma_m{\eta_-^{(|s|+1)}}$, there emerges a fermionic gauge symmetry generated by
$\hat{\epsilon}=\epsilon_+^{(|s|+1)}{\eta_+^{(|s|+1)}}$
\begin{eqnarray}
\label{fermion-ss}
\delta \psi_{\mu+}^{(|s|+1)}&=&\partial_{\mu}\epsilon_+^{(|s|+1)}
+i\alpha\sqrt{\frac{(1+|s|)}{2}}\,\epsilon_+^{(|s|+1)}\ ,
\nn \\
\delta \lambda_+^{(|s|+1,m)}&=&ig(1+|s|)\epsilon_+^{(|s|+1)}\ .
\end{eqnarray}
Using this gauge symmetry, one can eliminate $\psi_+^{(|s|+1)}$, such that the 10 by 10 mixing becomes 9 by 9 mixing. Diagonalising this system, we find that
the modes are determined by the partially-factorising
operator polynomial, of fifteenth order in $\pa$, given by
\be
\ell=|s|+1:\qquad \pa \Big(\Box- (|s|+2)\alpha^2\Big)^2 \tilde{{\cal O}}_5\ ,
\ee
where the explicit form of $\tilde{\cal O}_5$ can be deduced from the mixing equations. It is clear that is a massless spin-1/2 mode. Explicitly, it is a linear combination of $(\Lambda^{(|s|+1)}_+,\Psi^{(|s|+1)}_+,\Lambda^{(|s|+1)}_-,\Psi^{(|s|+1)}_-,\psi^{(|s|+1)}_-,$
$\chi^{(|s|+1)}_-, \chi^{(|s|+1)}_+,\lambda^{(|s|+1)}_+,\lambda^{(|s|+1)}_-)$ with mixing coefficients
\be
(8(1+2|s|),0,0,0,-2,-2i(1+2|s|),0,0,\sqrt{1+|s|}(1+4|s|).
\ee
\bigskip

\fbox{$\ell=|s|$}\\

The coupled system of linearized field equations for the spin-1/2 fields $(\Lambda_+^{(|s|)},\Psi_-^{(|s|)}, \psi_+^{(|s|)},\chi_+^{(|s|)},\lambda_-^{(|s|)})$ are
\bea
0&=&\pa \Lambda_+^{(|s|)}-\Psi_-^{(|s|)}-2\pa\psi_+^{(|s|)}
+2i\pa \chi_+^{(|s|)}-8gi\lambda_-^{(|s|)}\ ,
\\
0&=&i g |s| \Lambda_+^{(|s|)}+2g(1+|s|)i\psi_+^{(|s|)}+2g\chi_+^{(|s|)}+\pa\lambda_-^{(|s|)}=0,
\\
0&=&8gi\pa\lambda_-^{(|s|)}+\pa\Psi_-^{(|s|)}+2\Box\psi_+^{(|s|)}=0,
\\
0&=&(\Box+2\alpha^2-|s|\alpha^2)\pa\Lambda_+^{(|s| )}-(\Box+2\alpha^2-2|s|\alpha^2)\Psi_-^{(|s| )}+(2|s|-6)\alpha^2\pa\psi_+^{(|s| )}+6\alpha^2\,i\pa\chi_+^{(|s| )}\nn\\&&+32g|s|\alpha^2\,i\lambda_-^{(|s|) }\ ,
\\
0&=&\alpha^2\pa\Lambda_+^{(|s| )}-(1+|s|)\alpha^2\Psi_-^{(|s| )}-\pa(\Box+2\alpha^2+2|s|\alpha^2)\psi_+^{|s|}
-8g(1+|s|)\alpha^2\,i\lambda_-^{(|s|)}+2\alpha^2\,i\pa\chi_+^{(|s|)}\nn .
\\
\eea
Diagonalising this system, we find two massless modes which satisfy the relations
\be
\Psi_-^{(|s|)}=0\ ,\qquad \lambda_-^{(|s|)}=0\ ,\qquad |s|\Lambda_+^{(|s|)}-2i\chi_+^{(|s|)}+2(1+|s|)\psi_+^{(|s|)}=0\ .
\ee

\fbox{$\ell=|s|-1$}\\

When monopole charge $q\geq3$, there exist charge ``$-q/2$'' vector-spinor harmonics $\eta_m$ on $S^2$ possessing following properties
\be
D^m\left( _{s-\ft12}\eta^{(|s|-1)}\right)_m=0\ ,    \qquad    \sigma^m\left( _{s-\ft12}\eta^{(|s|-1)}\right)_m =0\ ,
\ee
where $s=(1-q)/2$. It follows that
\be
\sigma^m D_m \left( _{s-\ft12}\eta^{(|s|-1)}\right)_n=0\ .
\ee
Associated to this harmonics, there is a new spin-1/2 field $\tilde{\psi}^{(|s|-1)}$ satisfying
\be
\pa\Box\tilde{\psi}^{(|s|-1)}=0\ .
\ee

\subsection{Remarks on the non-supersymmetric spectrum}

   We saw in the supersymmetric vacuum that even at the $\ell=0$ level
in the Kaluza-Klein harmonic expansions, avoiding tachyons in the
four-dimensional spectrum required imposing the condition $M^2\ge
(5+2\sqrt{6})\alpha^2$ on the parameter $M$ in the six-dimensional
Lagrangian.  In the non-supersymmetric vacua we necessarily have
$M^2=\alpha^2$, and in fact having larger values for the background
monopole charge $q$ then the $q=\pm1$ supersymmetric case only
makes the tachyon problem worse, as can be seen from the vector mass operator
(\ref{vecmass0}).  For this reason, we shall not explore further
the precise details of the occurrence of tachyonic states in the
non-supersymmetric backgrounds.

   One feature of interest that we shall, however, comment on is the
occurrence of massless fermions in the non-supersymmetric backgrounds.
As can be seen from the spin-$\ft32$ operator in (\ref{gravitinomassless}),
there will be massless spin-$\ft32$ fields at level
$\ell=|s|=\ft12(q-1)$; thus these will occur in an $SU(2)$ multiplet
of dimension $2\ell+1 = q$.  At $\ell=|s|+1$, $\ell=|s|$ and
$\ell=|s|-1$ there will also be massless spin-$\ft12$ modes, as was discussed
in the spin-$\ft12$ section above.

\section{Conclusions}

  In this paper we have studied the complete linearised spectrum in
the $S^2$
Kaluza-Klein reduction of off-shell six-dimensional ${\cal N}=(1,0)$
gauged supergravity extended by a Riemann-squared superinvariant.
The higher-derivative terms in the six-dimensional theory can be expected
to imply the occurrence of ghosts.  As discussed in the introduction,
the usual argument for the positivity of the energies of states in a
supersymmetric background breaks down, and indeed we found that states
in the Kaluza-Klein spectrum could now have complex energies, thus
implying instabilities.

One way to understand the occurrence of complex masses is that there
are mixings between four-dimensional ghostlike and non-ghostlike modes.
This may be illustrated by the following simple example.  Consider
a set of fields $\phi_i$ with the Lagrangian
\be
{\cal L}= -\ft12 K_{ij}\, \del\phi_i\, \del\phi_j -
    \ft12 V_{ij}\phi_i\phi_j\,,
\ee
where $K_{ij}$ and $V_{ij}$ are constant symmetric matrices.  If the
eigenvalues of $K_{ij}$ are all positive, then $K_{ij}$ and $V_{ij}$
can be simultaneously diagonalised, by means of orthogonal transformations
combined with rescalings of the fields.  However, if $K_{ij}$ has
negative as well as positive eigenvalues, then the rescalings will
introduce factors of $\sqrt{-1}$ and the diagonalised mass matrix will
be complex.

  One way to avoid the ghost problems of the higher-derivative theory is
to treat it not as an exact model in its own right, but rather as an
effective theory valid at energy scales $\sqrt\Lambda$ much smaller than
$M$.  In this case the propagators are governed by the leading-order theory
without the higher-derivative terms, and these terms are treated as
interactions.  The Kaluza-Klein spectrum in the reduced
four-dimensional theory would then be simply that of the original
Salam-Sezgin model corresponding to the $M^2\rightarrow \infty$ limit. This spectrum has been given in \cite{Salam:1984cj} and our results agree in the $M^2\rightarrow \infty$ limit.
However, the Kaluza-Klein level number $\ell$ would have to be restricted
to lie below some maximum value, in order to satisfy the $\Lambda<<M^2$
limit.  Interestingly, this condition is sufficient to ensure that in the full, extended, theory,
the $m^2$ values of the retained modes would  all be real and positive.
This can be seen from (\ref{mul}), which indicates that the $m^2$
values will all be real and positive if $\ell$ is less than about
$M^2/(4\alpha^2)$.

A couple of remarks about the consistency of the Kaluza-Klein reduction are
in order.  Although we have restricted ourselves to a linearised analysis of
the four-dimensional spectrum, it should be emphasised that provided one
is keeping all the infinite towers of modes, then even at the full
non-linear order the reduction would still be consistent.  The truncation
of the spectrum at some maximum value of the level number $\ell$ that we
discussed in the previous paragraph would not, of course, be consistent
beyond the linear order, since the higher modes that were being set to zero
would be excited by sources involving the modes that are being retained.

Another more subtle question of consistency arises in this model also.  It
was shown in \cite{Gibbons:2003gp} that the Salam-Sezgin theory admits a non-trivial
consistent Pauli reduction on $S^2$, in which a finite subset of fields
including the $\ell=1$ triplet of Yang Mills gauge bosons are retained.
It would be interesting to see whether such a Pauli reduction is still
possible in the theory with the higher-derivative extension that we have
been considering in this paper.

Another interesting question is whether
the six dimensional model, with the auxiliary fields eliminated
in an order by order expansion in inverse powers of $M^2$, can be
embedded into the ten-dimensional heterotic string.  In the
gauged theory where $g\ne 0$, this continues to be a challenging problem
even before the higher-derivative terms are considered (although
some progress was made in a restricted sector of the theory in \cite{Cvetic:2003xr}).
For $g=0$, on the other hand, it was conjectured in \cite{Lu:2010ct} that
there is a relation with the 4-torus reduction of the heterotic theory
with Riemann-squared corrections that were constructed in \cite{Bergshoeff:1989de}.
This relation holds upon making a suitable truncation and performing an
S-duality transformation.  This conjecture was tested to lowest order
in the bosonic sector in \cite{Lu:2010ct}.

It would also be interesting to study exact solutions of the
higher-derivative six-dimensional supergravity.  While many solutions
of the Salam-Sezgin theory are known, exact solutions of the higher-derivative
theory, beyond the vacuum solutions we have discussed in this paper, are
scarce.  As far as we are aware, the only further example, which
exists only in the ungauged theory, is the self-dual string that was
found in \cite{lupang}.

A further question is whether there exist other quadratic-curvature
superinvariants over and above the Riemann-squared invariant of the
theory we have been considering.  This may have consequences for the
embedding of the theory in ten dimensions.

\section*{Acknowledgements}

We are very
grateful to F. Coomans and A. van Proeyen for many helpful discussions
on the six-dimensional model.
The research of C.N.P.~is supported in part by
DOE grant DE-FG03-95ER40917i, and that of E.S. is supported by in part
by NSF grant PHY-0906222.

\newpage

\appendix

\section{Conventions}

We choose the 6$D$ gamma matrices to be
\begin{eqnarray}
  && {\Gamma}^0=\gamma^0\otimes\sigma_3\ ,\qquad {\Gamma}^1=\gamma^1\otimes\sigma_3\ ,\qquad {\Gamma}^2=\gamma^2\otimes\sigma_3\ ,
\nonumber\\
  &&{\Gamma}^3=\gamma_3\otimes\sigma_3\ ,\qquad {\Gamma}^4=1_{4\times4}\otimes\sigma_1\ ,
  \qquad {\Gamma}^5=1_{4\times4}\otimes\sigma_2\ .
\end{eqnarray}
One can check that
\begin{eqnarray}
&& {\Gamma}^0{\Gamma}^{\mu\dagger}{\Gamma}^0={\Gamma}^{\mu}\ ,\qquad B={\Gamma}^3\Hat{\Gamma}^5\ ,
\nonumber\\
  && B^*B=-1\ ,\qquad B{\Gamma}^{\mu}B^{-1}={\Gamma}^{\mu*}\ .
\end{eqnarray}
The $SU$(2) symplectic-Majorana-Weyl spinor is defined by
\begin{equation}\label{}
    \psi^{*i}=(\psi_{i})^*=\epsilon^{ij}B\psi_{j}.
\end{equation}
A useful formula related to the $SU$(2) symplectic-Majorana-Weyl spinor is
\begin{equation}\label{}
    \bar{\lambda}^i{\Gamma}^{(n)}\psi^j=t_n\bar{\psi}^j{\Gamma}^{(n)}\lambda^i\ ,\qquad t_{n}=\left\{
  \begin{array}{ll}
    +, & n=1,2,5,6; \\
    -, & n=0,3,4.
  \end{array}
\right.
\end{equation}

\section{Spin-weighted Harmonics on $S^2$}

  In this appendix, we give an elementary construction of the spin-weighted
spherical harmonics.  This is based on
a specialisation of results for the analogous
harmonics in the complex projective space $CP^n$, which were discussed in
\cite{homapo}.  Since the azimuthal label $m$ on the spin-weighted
spherical harmonics $_sY_{\ell m}$ plays an important role in the
derivations in this appendix, we shall suspend our convention used in the
body of the paper of suppressing the $m$ label.  In order to avoid confusion
with coordinate indices, we shall use $i$, $j$,... for coordinate indices
on $S^2$ in this appendix.

\subsection{Scalar spin-weighted harmonics}

   The scalar
spin-weighted spherical harmonics $_sY_{\ell m}$ are the eigenfunctions
of the {\it charged} scalar Laplacian $\Box_{(s)}$ on the unit $S^2$,
carrying electric charge $s$, in the
presence of a Dirac monopole with potential $A=-\cos\theta\, d\phi$:
\be
\Box_{(s)} \equiv \fft1{\sin\theta}\, \fft{\del}{\del\theta}
 \Big( \sin\theta\, \fft{\del}{\del\theta}\Big) +
\fft{1}{\sin^2\theta}\,
\Big(\fft{\del}{\del\phi} + \im\, s\, \cos\theta\Big)^2\,.
\label{boxs}
\ee
In the language of differential forms, the charged Laplacian operator
on the unit $S^2$ with metric $d\Omega_2^2=d\theta^2 + \sin^2\theta\, d\phi^2$
may be written in terms of the charged Hodge-de Rham operator
\be
\Delta \equiv {*D*}D + D{*D*}\,,\label{HdR}
\ee
as
\be
-\Box_{(s)} =\Delta \,,
\ee
where $D$ is the charge-$s$ gauge-covariant exterior derivative
\be
D= d + \im s \cos\theta d\phi\,.\label{Ddef}
\ee

The spin-weighted harmonics
may be constructed by starting with the four-dimensional
scalar Laplacian on $\Bbb{C}^2$, and then embedding the unit $S^3$, viewed as
a $U(1)$ bundle over $S^2$, in $\Bbb{C}^2$.    Introducing complex coordinates
$Z^a$ on $\Bbb{C}^2$, the four-dimensional Laplacian is
\be
\Box_4 = 4 \fft{\del^2}{\del Z^a \del\bar Z_a}\,.
\ee
Clearly, if we define functions
\be
f= T_{a_1\cdots a_p}{}^{b_1\cdots b_q}\, Z^{a_1}\cdots Z^{a_p}\,
\bar Z_{b_1}\cdots \bar Z_{b_q}\,,\label{TZZ}
\ee
where $T_{a_1\cdots a_p}{}^{b_1\cdots b_q}$ is symmetric in its upper
and its lower indices, and traceless with respect to any contraction of upper
and lower indices, then they will satisfy
\be
\Box_4 f=0\,.\label{4boxf}
\ee

   Writing
\be
Z^1= r\,e^{\im(\psi+\phi)/2}\, \cos\ft12\theta\,,\qquad
Z^2= r\,e^{\im(\psi-\phi)/2}\, \sin\ft12\theta\,,\label{Zdef}
\ee
the Euclidean metric on $\Bbb{C}^2$ is expressible as
\be
ds_4^2 = dr^2 + r^2\, d\Omega_3^2\,,
\ee
where
\be
d\Omega_3^2 = \ft14 (d\psi+\cos\theta\, d\phi)^2 + \ft14(d\theta^2 +
\sin^2\theta\, d\phi^2)\label{S3met}
\ee
is the metric on the unit 3-sphere.  The four-dimensional Laplacian
is given by
\be
\Box_4 = \fft1{r^3}\, \fft{\del}{\del r}\Big(r^3\, \fft{\del}{\del r}\Big) +
\fft1{r^2}\, \Box_3,\label{4box3box}
\ee
where $\Box_3$ is the Laplacian on the unit $S^3$.  Noting from (\ref{TZZ})
and (\ref{Zdef}) that $f$ takes the form
\be
f = r^{p+q}\, e^{\im(p-q)\psi/2}\, Y(\theta,\phi)\,,\label{fZZ}
\ee
then  (\ref{4boxf}) and (\ref{4box3box}) imply that
\be
\Box_3 \Big(e^{\im(p-q)\psi/2}\, Y(\theta,\phi)\Big) =
-(p+q)(p+q+2)\, e^{\im(p-q)\psi/2}\, Y(\theta,\phi)\,.\label{3boxY}
\ee
 From (\ref{S3met}), $\Box_3$ is given by
\be
\ft14\,\Box_3 =  \fft1{\sin\theta}\, \fft{\del}{\del\theta}
 \Big( \sin\theta\, \fft{\del}{\del\theta}\Big) +
\fft{1}{\sin^2\theta}\,
\Big(\fft{\del}{\del\phi} - \cos\theta\,\fft{\del}{\del\psi}\Big)^2
   +\fft{\del^2}{\del\psi^2}\,,
\ee
and so if we define
\be
p=\ell-s\,,\qquad q=\ell + s\,,\label{pqdef}
\ee
then (\ref{3boxY}) implies that $Y(\theta,\phi)$ satisfies
\be
-\Box_{(s)}\, Y = [\ell(\ell+1) -s^2]\, Y\,,
\ee
where $\Box_{(s)}$ is the charged scalar Laplacian on $S^2$ that we defined in
equation (\ref{boxs}).  Up to an overall conventional normalisation, we
see that $Y(\theta,\phi)$ constructed from (\ref{TZZ}) and (\ref{fZZ})
is nothing but a spin-weighted spherical harmonic $_sY_{\ell m}$.  Since
$p$ and $q$ in (\ref{TZZ}) are non-negative integers, and they are related to
$\ell$ and $s$ by (\ref{pqdef}), it follows that
\be
\ell \ge |s|\,.
\ee
It is easily seen that the number of independent traceless symmetric tensors
$T_{a_1\cdots a_p}{}^{b_1\cdots b_q}$ in (\ref{TZZ}) is equal to
$1+p+q$, and hence we have constructed the $2\ell+1$ spin-weighted
spherical harmonics $_sY_{\ell m}$ at level $\ell$ satisfying
\be
-\Box_{(s)}\, _sY_{\ell m} = [\ell(\ell+1) -s^2]\, _sY_{\ell m}\,,\qquad
\ell\ge |s|\,,\qquad -\ell \le m\le \ell\,.\label{scalarevs}
\ee
Note that $s$, $\ell$ and $m$ are either all integers, or else all
half-integers.

   With the conventional normalisation, the spin-weighted spherical
harmonics satisfy the relations
\bea
{\cal D}_{\!-} \,\, _sY_{\ell m}\equiv
\Big(\fft{\del}{\del\theta} + m\csc\theta + s \cot\theta\Big) \, _sY_{\ell m}
&=&
\sqrt{(\ell+s)(\ell+1-s)}\,\, _{s-1}Y_{\ell m}\,,\nn\\
{\cal D}_{\!+}\, \,_{s-1}Y_{\ell m}\equiv
\Big(\fft{\del}{\del\theta} - m\csc\theta - (s-1) \cot\theta\Big) \,
_{s-1}Y_{\ell m} &=&
-\sqrt{(\ell+s)(\ell+1-s)}\,\, _sY_{\ell m}\,.\label{sYlm}
\eea

\subsection{Vector spin-weighted harmonics}

   The spin-weighted vector harmonics are the eigenfunctions of the
charged Hodge-de Rham operator (\ref{HdR}) acting on 1-forms:
\be
\Delta V = \tilde\lambda\, V \ , \qquad V=dy^i V_i\ .
\label{Vev}
\ee
Generically, these eigenfunctions can be constructed from the scalar
spin-weighted harmonics $_sY_{\ell m}$ (denoted simply as $Y$ below) by writing
\be
V = DY + \mu\, {*D}Y\,,
\ee
where $D=d + \im\, s \cos\theta\, d\phi$ is the gauge-covariant exterior
derivative.
We shall write the eigenvalues for the scalar spin-weighted harmonics,
given by (\ref{scalarevs}), simply as $\lambda$, so that
\be
\Delta Y =\lambda Y\,,\qquad \lambda= \ell(\ell+1) -s^2,.
\ee
Noting that
\be
D^2= -\im s \Omega_2\,,
\ee
where $\Omega_2=\sin\theta\, d\theta\wedge d\phi$ is the volume form on
the unit $S^2$, that $D{*D*} (DY)= D\Delta Y= \lambda DY$,
${*D*}D(DY) = -\im s \, {*D}({*\Omega_2} Y) =-\im s\, {*D}Y$ and that
${\Delta *} = {*\Delta}$,
we see that
\be
\Delta V =(\lambda+ \im \mu s)DY + (\mu\lambda-\im s){*DY}\,.
\ee
Thus $V$ is an eigenfunction, satisfying (\ref{Vev}), if $\mu=\pm \im$,
and so generically we get two distinct vector eigenfunctions $V^\pm$
with corresponding
eigenvalues $\tilde\lambda_\pm$ from each scalar eigenfunction $Y$, where
\be
V^\pm = DY \mp \im {*DY}\,,\qquad \tilde\lambda_\pm = \lambda\pm s\,.
\label{tildelam}
\ee
In terms of $\ell$ and $s$, these eigenvalues are given by
\be
\tilde\lambda_+ = (\ell+s)(\ell+1-s)\,,\qquad
\tilde\lambda_- = (\ell-s)(\ell+1+s)\,.
\ee
Note that the vector harmonics $V^\pm$ obey the complex duality conditions
\be
{* V^\pm}= \pm \im V^\pm\,.
\ee

  A special case arises if the scalar eigenvalue $\lambda$ is equal to $s$
or $-s$.
(Since $\lambda$
is necessarily non-negative, the former can only arise if $s$ is positive, and
it implies $\ell=s$, while  the latter arises if $s$ is negative, and implies
$\ell=-s$.)  Calculating the norm of $V^-$, we find
\bea
\int{*{\bar V^-}}\wedge V^- &=&\int({*D}\bar Y -\im {*D}\bar Y )\wedge
  (DY +\im\, {*D}Y)= 2\int({*D}Y\wedge D\bar Y - \im\, DY\wedge D\bar Y)\nn\\
&=& 2\int( (D{*D} Y)\bar Y - \im\, (D^2 Y)\bar Y)
= 2(\lambda -s) \int |Y|^2\,,
\eea
and so $V^-=0$ if $\lambda=s$.  A similar calculation shows $V^+=0$ if
$\lambda=-s$.  Thus if $\lambda=s$ then the mode $V^-$, which from
(\ref{tildelam}) would have had eigenvalue $\tilde\lambda_-=0$, is absent.
Similarly, if $\lambda=-s$ then $V^+$, which would likewise have had
eigenvalue $\tilde\lambda_+=0$, is absent.

  In fact vector spin-weighted zero modes of $\Delta$ {\it do} arise, but
they cannot be constructed from scalar harmonics in the manner described above.
If $\Delta V=0$ then integrating ${*{\bar V}} \Delta V$ over the sphere
implies
\be
\int(|D{*V}|^2 + |DV|^2)=0
\ee
and hence $V$ is (gauge) closed and co-closed,
\be
DV=0\,,\qquad D{*V}=0\,.\label{DV}
\ee
We can project into the self-dual and
anti-self-dual subspaces, and thus seek 1-forms $V$ satisfying
\be
{*V}= \pm \im V\,,\qquad DV=0\,.
\ee
Making the ansatz
\be
V= e^{\im m\phi}\, (fd\theta + g d\phi)\,,
\ee
where $f$ and $g$ are functions of $\theta$,
we find ${*V}=e^{\im m\phi}\, (-f\sin\theta\, d\phi +
   g\, \csc\theta\, d\theta)$ and hence the duality condition implies
\be
g=\pm\im\, f\sin\theta\,.
\ee
The condition $DV=0$ implies $g'= \im f(m + s \cos\theta)$, and hence
we obtain
\be
f = c (\sin\theta)^{-1\pm s}\, (\tan\ft12\theta)^{\pm m}\,.
\ee
Thus if $s\ge 1$
we obtain regular self-dual harmonics (${*V^+}= +\im\, V^+$) given by
\be
V^+= (\sin\ft12\theta)^{s-1+m}\, (\cos\ft12\theta)^{s-1-m}\,
  e^{\im m\phi}\, (d\theta+ \im\, \sin\theta\, d\phi)\,, \qquad
  -(s-1)\le m  \le (s-1)\,,\label{selfdual}
\ee
while if $s\le -1$ we obtain regular anti-self-dual harmonics (${*V^-}=-
\im\, V^-$)
given by
\be
V^-= (\sin\ft12\theta)^{-s-1-m}\, (\cos\ft12\theta)^{-s-1-m}\,
  e^{\im m\phi}\, (d\theta - \im\, \sin\theta\, d\phi)\,, \qquad
  s+1\le m  \le -s-1\,.\label{asd}
\ee
In each case, these charge-$s$ vector harmonics form an $\ell= |s|-1$
representation of $SU(2)$, as evidenced by the $(2|s|-1)$-fold multiplet
of $m$ values.

  In summary, each spin-weighted scalar harmonic $Y$ with eigenvalue
$\lambda=\ell(\ell+1)-s^2$ and with with $\ell\ge |s|+1$
gives rise to two spin-weighted vector harmonics, namely a
self-dual harmonic $V^+$ with eigenvalue $\tilde\lambda_+=(\ell+s)(\ell+1-s)$
for the charge$-s$ Hodge-de Rham operator $\Delta$, and an anti-self dual
harmonic $V^-$ with
eigenvalue $\tilde\lambda_- =(\ell-s)(\ell+1+s)$.
However, the lowest-level spin-weighted scalar harmonic, with $\ell=|s|$,
gives rise to only one spin-weighted vector harmonic, namely $V^+$ if
$s$ is positive, or $V^-$ if $s$ is negative.  The ``missing'' vector
harmonic when $\ell=|s|$ would have been a zero-mode of $\Delta$.  In its
place, a zero-mode harmonic satisfying $\Delta V=0$ {\it does} occur,
but it cannot be constructed from the scalar spin-weighted harmonics.
It corresponds to $\ell  =|s|-1$, and therefore has multiplicity $2|s|-1$.

\subsection{Spin-$\ft12$ spin-weighted harmonics}

    We may define the spin-weighted spinor harmonics to be charged
solutions of the Dirac equation in the monopole background.  They may, in
general, be constructed from the scalar spin-weighted harmonics, as we now
describe.  We first note that there exist two charged gauge-covariantly
constant spinors on $S^2$ with the monopole background, satisfying $D\eta=0$
where we now add a spin connection term to the gauge-covariant
exterior derivative,
\be
D= \nabla + \im\, s\, \cos\theta\, d\phi\,,\qquad \nabla\equiv d +
\ft14 \omega_{ab}\sigma^{ab}\,
\ee
and, when acting on $\eta^\pm$,
$s=\pm\ft12$.  This can be seen from the integrability condition
\be
0=[D_i,D_j]\eta= \ft14 R_{ijk\ell}\sigma^{k\ell}\eta - \im\, s\, \ep_{ij}\,
  \eta= \im\, \ep_{ij}\, (\ft12\sigma_3 -s)\eta\,
\ee
from which we see that there exist two solutions:
\bea
s=\ft12:&& D\eta_+=0\,,\qquad \sigma_3\eta_+=\eta_+\,,\nn\\
s=-\ft12:&& D\eta_-=0\,,\qquad \sigma_3\eta_-= -\eta_-\,.
\eea
Using the standard basis for the Pauli matrices $\sigma_i$, the solutions,
we find that the gauge-covariantly constant spinors $\eta_\pm$ are given
by
\be
\eta_+=\begin{pmatrix} 1 \\ 0\end{pmatrix} \,,\qquad
\eta_-=\begin{pmatrix} 0 \\ 1\end{pmatrix}\,.
\ee
From these spinors, which are normalised so that $\bar\eta_+\eta_+=
\bar\eta_-\eta_-=1$, we may construct the gauge-covariantly constant vector
\be
U = \bar\eta_- \sigma^i\eta_+\, \del_i =  \fft{\del}{\del\theta} +
  \im\csc\theta\, \fft{\del}{\del\phi}\,,\label{Udef}
\ee
which has charge $s=1$.  Its complex conjugate $\bar U=\del/\del\theta
-\im\,\csc\theta\, \del/\del\phi$ has charge $s=-1$.  In fact $U$ is the
holormorphic $(1,0)$-form on the K\"ahler manifold $S^2$, satisfying
$J_i{}^j\, U_j= \im\, U_i$, where $J_{ij}=\ep_{ij}$ is the K\"ahler form.
Note that
\be
\sigma^i\eta_+ = U^i\, \eta_-\,,\qquad \sigma^i\eta_-= \bar U^i\, \eta_+\,.
\label{sigeta}
\ee
The operators $U^i D_i$ and $\bar U^i D_i$ give precisely
${\cal D}_+$ and ${\cal D}_-$, defined in (\ref{sYlm}), when acting on
$_sY_{\ell m}$ and $_{s-1}Y_{\ell m}$ respectively.  It is now clear why
${\cal D}_+$ and ${\cal D}_-$ raise and lower the charge of the spin-weighted
scalar harmonics by one unit, since $U^i$ and $\bar U^i$ are gauge-covariantly
constant vectors carrying $+1$ and $-1$ charge respectively.

   The solutions of the charged Dirac equation can be expressed in terms of
chiral spinors $\psi_+$ and anti-chiral spinors $\psi_-$, satisfying
\be
\sigma^i D_i \psi_+ = \im\, \lambda_+\, \psi_-\,,\qquad
\sigma^i D_i \psi_- = \im\, \lambda_-\, \psi_+\,.\label{Diracpm}
\ee
Since $(\sigma^i D_i)^2\psi = D^iD_i \psi +(\tilde s\sigma_3-\ft12)\psi$ where
$\psi$ is any spinor with charge $\tilde s$, we have
\be
-D^iD_i\psi_+ = (\lambda_+\lambda_- + \tilde s-\ft12)\psi_+\,,\qquad
-D^iD_i\psi_- = (\lambda_+\lambda_- - \tilde s-\ft12)\psi_-\,.
\ee
The product $\lambda_+\lambda_-$ is therefore uniquely determined, as a
function of $\tilde s$ and $\ell$, for each spinor eigenfunction of the
second-order operator $D^iD_i$.  The values of $\lambda_+$ and $\lambda_-$
separately are not determined, but depend upon the choice of relative
normalisation for $\psi_+$ and $\psi_-$ in (\ref{Diracpm}).

   It is convenient to consider spinor eigenfunctions $\psi_+$ and $\psi_-$
with charge $\tilde s=s-\ft12$.  We may construct these from the scalar
spin-weighted harmonics $_sY_{\ell m}$ discussed earlier by writing
\be
\psi_- = \eta_- \left(_sY_{\ell m}\right)\ , \qquad \psi_+ = \eta_+\left(_{s-1}Y_{\ell m}\right)\ .
\label{psipm}
\ee
Note these $\psi_\pm$ are denoted by ${\tilde\eta}^{(\ell)}_\pm$ in \eq{nota1}. For brevity in notation, however,  we shall
continue to use the notation $\psi_\pm$ instead in this section. Acting on these with the Dirac operator, we find, using (\ref{sYlm}),
(\ref{Udef}) and (\ref{sigeta}), that
\bea
\sigma^iD_i\psi_- &=& \sigma^i\eta_-\, D_i \, _sY_{\ell m} =
\eta_+\, \bar U^i D_i\, _sY_{\ell m} \nn\\
&=& \eta_+\,
\Big(\fft{\del}{\del\theta} + m\csc\theta + s \cot\theta\Big) \, _sY_{\ell m}
=\eta_+\, \sqrt{(\ell+s)(\ell+1-s)}\, _{s-1}Y_{\ell m}\,,\nn\\
\sigma^iD_i\psi_+ &=& \sigma^i\eta_+\, D_i \, _{s-1}Y_{\ell m} =
\eta_-\, U^i D_i\, _{s-1}Y_{\ell m} \\
&=& \eta_-\,
\Big(\fft{\del}{\del\theta} - m\csc\theta - (s-1) \cot\theta\Big)
\, _{s-1}Y_{\ell m}
=-\eta_-\, \sqrt{(\ell+s)(\ell+1-s)}\, _sY_{\ell m}\,,\nn
\eea
and hence
\be
\sigma^iD_i \psi_- =  \sqrt{(\ell+s)(\ell+1-s)}\,\psi_+\,,\qquad\
\sigma^iD_i \psi_+ = - \sqrt{(\ell+s)(\ell+1-s)}\,\psi_-\,.
\ee

   It is worth remarking that there is an alternative procedure
that in general constructs the charged spin-$\ft12$ harmonics from
scalar harmonics, in which only one of the gauge-covariantly constant
spinors is required.  For example, using only $\eta_-$ we can construct
the negative-chirality spin-$\ft12$ harmonics $\psi_-$ as in the first
equation in (\ref{psipm}), while for the positive-chirality harmonics
we take
\be
\psi_+' = \sigma^i\eta_- \, D_i \,_sY_{\ell m}\,.\label{psiprime}
\ee
A straightforward calculation shows that
\be
\sigma^i D_i\, \psi_+' = -(\ell+s)(\ell+1-s)\, \eta_-\, _sY_{\ell m}\,.
\ee
The harmonics $\psi_+'$ are in general proportional to the harmonics
$\psi_+$ given in (\ref{psipm}).  However, the charge $\ft12$ harmonic
$\psi_+=\eta_+$ itself (which is a zero mode of the Dirac operator)
cannot be constructed using (\ref{psiprime}), since
it would require taking $s=1$ and $\ell=0$, for which $_sY_{\ell m}$ does
not exist.

   It might also seem that charge $-s-\ft12$
zero modes $\psi_+'$ would be obtained if
$s$ were negative and $\ell=-s$.  However,
calculating the norm of $\psi_+'$, we find
\be
\int_{S^2} |\psi_+'|^2 \sqrt{g} d^2x = (\ell+s)(\ell+1-s)\,
\int_{S^2} |_sY_{\ell m}|^2\, \sqrt{g} d^2x\,,
\ee
and thus $\psi_+'$ would actually be identically zero if $\ell=-s$.  These
putative zero modes are in fact not obtained by the construction for $\psi_+$
in (\ref{psipm}) either, since this would require the use of
scalar harmonics
with $\ell$ smaller than the magnitude of their spin weight.

\subsection{Spin-$\ft32$ spin-weighted harmonics}

   The general spin-$\ft32$ harmonics $\eta_i$ can be decomposed into chiral
and antichiral projections $\eta_i^\pm$ satisfying
\be
\sigma^iD_i\, \eta^+_j = \lambda_+\, \eta_j^-\,,\qquad
\sigma^iD_i\, \eta^-_j = \lambda_-\, \eta_j^+\,.
\ee
Each chiral projection admits a decomposition of the form
\be
\eta_i^\pm  = \sigma_i \psi^\mp + \eta_{\{i\}}^\pm + \tilde\eta_i^\pm\,,
\ee
where $\eta_{\{i\}}^\pm$ is longitudinal and gamma traceless,
$\sigma^i\, \eta_{\{i\}}^\pm =0$, and $\tilde\eta_i^\pm$ is transverse
and gamma traceless, satisfying $D^i\tilde\eta_i^\pm=0$ and
$\sigma^i\, \tilde\eta_i^\pm=0$.  We can write $\eta_{\{i\}}^\pm$ in
terms of spin-$\ft12$ modes $\eta^\pm$ as
\be
\eta_{\{i\}}^\pm = 2D_i\psi^\pm - \sigma_i \sigma^j D_j\psi^\pm
  = (D_i \mp\im\, \ep_i{}^j\, D_j)\psi^\pm\,.
\ee
In fact $\eta_{\{i\}}^\pm$ can alternatively be written in terms of the vector harmonics $V^\pm$ constructed
from scalar harmonics as in (\ref{tildelam}), by taking
\be
\ \eta^\pm_{\{i\}} = V^\pm_i\, \eta^{\pm}\,.
\label{fromvec1}
\ee
The gamma-tracelessness of $\eta^\pm_{\{i\}}$ follows immediately from the fact that $V^\pm_i$ and
$\sigma^i\eta^\pm$ are either both self-dual or both anti-self dual.  The
charge carried by $\eta^\pm_{\{i\}}$ will,
of course, be equal to $s\pm\ft12$, where $s$ is the charge of $V_i^\pm$.

The transverse traceless spin-$\ft32$ harmonics $\tilde\eta_i$ can be constructed in the same way, and
are given by (\ref{fromvec1}) except that now, $V^\pm_i$ are the self-dual vector harmonics (\ref{selfdual})
or the anti-self dual harmonics (\ref{asd}) that cannot be constructed from scalar harmonics.  Since such
$V^\pm_i$ vectors arise only when $s\ge1$ or $s\le -1$ respectively, the transverse traceless spin-$\ft32$
harmonics $\tilde\eta^\pm_i$ arise only for charges $\tilde s\ge\ft32$ or $\tilde s\le -\ft32$ respectively.

All necessary properties of the spin-$\ft32$ harmonics follow from the
properties of the lower-spin harmonics that we discussed previously.

\section{Spin Projection Operators}

The well known spin projector operators associated with a second rank symmetric tensor field are given by
\cite{VanNieuwenhuizen:1981ae}
\bea
&&\cP_{\mu\nu,\rho\sigma}^{2}=\ft12(\theta_{\mu\rho}\theta_{\nu\sigma}
+\theta_{\mu\sigma}\theta_{\nu\rho}-\ft23\theta_{\mu\nu}\theta_{\rho\sigma})\ ,
\nn\w2
&&\cP_{\mu\nu,\rho\sigma}^{1}=\ft12(\theta_{\mu\rho}\omega_{\nu\sigma}
+\theta_{\mu\sigma}\omega_{\nu\rho}+\theta_{\nu\rho}\omega_{\mu\sigma}+\theta_{\nu\sigma}\omega_{\mu\rho})\ ,
\nn\w2
&&\cP_{\mu\nu,\rho\sigma}^{(0,s)}=\ft13\theta_{\mu\nu}\theta_{\rho\sigma}\ ,
\nn\w2
&&\cP_{\mu\nu,\rho\sigma}^{(0,\omega)}=\omega_{\mu\nu}\omega_{\rho\sigma}\ ,
\eea
where
\be
\theta_{\mu\nu}=\eta_{\mu\nu}-\Box^{-1}\partial_{\mu}\partial_{\nu}\ ,\qquad \omega_{\mu\nu}=\Box^{-1}\partial_{\mu}\partial_{\nu} \ .
\ee
Similarly, the spin projector operators associated with vector-spinor field take the form
\bea
P_{\mu\nu}^{3/2} &=& \theta_{\mu\nu}-\frac13 \theta_\mu\theta_\nu\ ,
\nn\w2
(P_{11}^{1/2})_{\mu\nu} &=& \frac13 \theta_\mu\theta_\nu\ ,\qquad
(P_{12}^{1/2})_{\mu\nu}  = \frac1{\sqrt 3} \theta_\mu\omega_\nu\ ,
\nn\w2
(P_{21}^{1/2})_{\mu\nu}  &=& \frac1{\sqrt 3} \omega_\mu\theta_\nu\ ,\qquad
(P_{22}^{1/2})_{\mu\nu} =\omega_\mu\omega_\nu\ ,
\eea
where
\be
\theta_\mu = \theta_{\mu\nu} \gamma^\nu\ ,\qquad \omega_\mu= \omega_{\mu\nu}\gamma^\nu\ .
\ee

\end{document}